\documentclass{aa}
\usepackage{graphicx}
\usepackage{txfonts}
\usepackage{pifont}
\usepackage{bigstrut}
\usepackage{lscape}
\usepackage{rotating}
\usepackage{subfigure}
\usepackage{color}

\begin{document} 

   \title{Synthetic photometry for carbon-rich giants}
      
   \subtitle{V. Effects of grain-size-dependent dust opacities}

   \author{K. Eriksson
          \inst{1}
          \and
          S. H\"{o}fner \inst{1}
          \and
          B. Aringer\inst{2}\fnmsep\thanks{Table C.1 is only available in electronic form at the CDS via
          anonymous ftp to cdsarc.cds.unistra.fr (130.79.128.5) or via https://cdsarc.cds.unistra.fr/cgi-bin/qcat?J/A+A/.}
          }

   \institute{Theoretical Astrophysics, Department of Physics and Astronomy, Uppsala University, Box~516,
        SE-75120 Uppsala
        \and
   Department of Astrophysics, University of Vienna,
              T\"urkenschanzstrasse 17, A-1180 Vienna\\             }

   \date{Received ; accepted }

\abstract
{The properties and the evolution of asymptotic giant branch (AGB) stars are
strongly influenced by their mass loss through a stellar wind.
This, in turn, is believed to be caused by radiation pressure due to the absorption and 
scattering of the stellar radiation by the dust grains formed in the atmosphere. 
The optical properties of dust are often estimated using the small particle limit (SPL) 
approximation, and it has been used frequently in modelling AGB stellar winds when 
performing radiation-hydrodynamics (RHD) simulations.} 
{We aim to investigate the
effects of replacing the SPL approximation by detailed Mie calculations of the size-dependent
opacities for grains of amorphous carbon forming in C-rich AGB star atmospheres.
}
{We performed RHD simulations for a large grid of carbon star atmosphere+wind models 
with different effective temperatures, luminosities, stellar masses, carbon excesses, and 
pulsation properties. 
Also, {\it \emph{a posteriori}} radiative transfer calculations for many radial structures 
(snapshots) of these models were done, resulting in spectra and filter magnitudes.
}
{We find that, when giving up the SPL approximation, the wind models become more strongly
variable and more dominated by gusts, although the average mass-loss rates and outflow speeds 
do not change significantly; the increased radiative pressure on the dust throughout its formation 
zone does, however, result in smaller grains and lower condensation fractions (and thus higher 
gas-to-dust ratios). 
The photometric $K$ magnitudes are generally brighter, but at $V$ the effects of using 
size-dependent dust opacities are more complex: brighter for low mass-loss rates and 
dimmer for massive stellar winds. 
}
{Given the large effects on spectra and photometric properties, it is necessary to use the 
detailed dust optical
data instead of the simple SPL approximation in stellar atmosphere+wind modelling 
where dust is formed.
}



\maketitle
%

\section{Introduction}
Asymptotic giant branch (AGB) stars are in complex, very late evolutionary stages 
as descendents of stars that started their lives on the main sequence with masses 
of $\sim$ 0.8 --  8 M$_\odot$.
They mainly generate energy in a H-burning shell, interrupted by periods of intense 
He-burning in an inner shell during the so-called He-shell-flashes (see, e.g. \citet{Herw05} 
for details).
Many of the AGB stars  are variable in brightness (as Mira stars or other types of 
long-period variables, LPVs) -- this has often been interpreted as pulsations, 
especially for Miras.
Another important point is the chemical evolution of the stellar atmosphere when the products of 
nucleo\-synthesis are dredged-up to the surface through the deep convection zone; 
in particular, the carbon produced by He burning may turn an M-type AGB star into a 
carbon star if/when the atmospheric abundance of carbon exceeds that of oxygen.  

The evolution of stars in the AGB stage is more determined by mass loss 
from the surface than by the growth of their cores.
The mass loss from AGB stars has been the subject of research for at least the past 
60 years, both observationally and theoretically (see, e.g. the review by \citet{HoOl18}). 
One popular mass-loss scenario is the pulsation-enhanced dust-driven outflow, 
where the stellar pulsations generate outward shocks that carry  material to 
higher and cooler layers where dust can condense;
the opacity of the dust causes the material to experience a radiation pressure that can, 
under suitable conditions, induce and sustain a stellar wind.

This scenario has been modelled theoretically with the radiation-hydrodynamics code
DARWIN \citep[see][and below]{hoefetal16} 
and resulting synthetic observables have been presented in a series of papers.
In the first article in this series, Paper I, \citet{AGNML09} calculated 
spectra and photometric indices for hydrostatic carbon star photospheric models. 
\citet{NoAHL11}, Paper II, found that including a pulsating atmosphere and dust formation 
in the wind produced observables largely in agreement with those of the carbon star RU~Vir. 
In Paper III, \citet{NoAHE13} presented a sequence of carbon star atmosphere and wind 
models with varying parameters to illustrate the properties of early to late carbon stars.
In the previous article in this series, \citet{E+14}, Paper IV, we calculated 
synthetic colours and spectra for a large grid of solar-metallicity carbon star 
atmosphere and wind models. The mass-loss rates of these models, together
with other DARWIN results presented by \citet{MatWH10} and \citet{Bladh+19}, were used
by \citet{Pastorelli+19, Pastorelli+20} in their population synthesis studies of SMC and LMC
AGB stars, demonstrating the critical role of a realistic mass-loss description.

Photometric observational data, both in the form of long time series for individual AGB stars
\citep[e.g.][]{W+06,Menz+06} and large surveys 
\citep[e.g.][]{Sosz+09,Groenw+20}, has played an important role in studying the stellar 
pulsations, and their effects on the dust-forming atmospheres. 
While high-angular-resolution techniques have made it possible to resolve the 
atmospheres and wind formation zones of a handful of nearby AGB stars 
\citep[e.g.][]{Sacu+11,Khou+16,Witt+17}, photometry is still a critical tool for studying 
large samples with different stellar parameters.
As these stars are very luminous they can be observed also in other galaxies.
In addition, large spectroscopic surveys, combined with efficient spectral
classification methods, can contribute to our understanding of evolved stars
\citep[e.g.][]{Li+18}.

In the present paper, we improve on our earlier calculations, using dust opacities 
that depend on the size of the dust grains, in contrast to Paper IV where we assumed 
the small particle limit (SPL) for the dust opacity.
The current study was motivated by our previous results, showing that grains tend to 
reach sizes beyond the regime where the SPL applies.
We find that the mass-loss rates and outflow velocities are similar to the 
previous results, but the dust condensation (and thus the dust-to-gas ratio) is much smaller, 
and the photometric filter magnitudes and colours are also quite strongly affected.


\section{Dynamic models and radiative transfer}
\label{s:modelling}

The modelling of dynamic structures and synthetic observables follows the general
approach described in Papers II -- IV of this series. 
Here, we only give a brief summary, referring 
the reader to these and other papers for a more comprehensive description. 
A more detailed description of the size-dependent dust opacities used here, 
in contrast to Papers II -- IV, is given below.

\subsection{Dynamic atmosphere and wind models}
\label{s:dynmodelling}

The time-dependent radial structures of atmospheres and winds are produced by simultaneously
solving the equations of hydrodynamics, frequency-dependent radiative transfer, 
and non-equilibrium dust formation (describing nucleation and growth of amorphous 
carbon grains) using the DARWIN code \citep[see][]{HoGAJ03, hoefetal16}.
Each dynamic model is represented by a sequence of snapshots of radial structures, 
typically covering hundreds of pulsation periods.
Wind properties are derived from the time sequences as described in Sect. \ref{s:DMAdyndust}.

The computation of a dynamical model starts from a hydrostatic, dust-free configuration 
characterized by the fundamental stellar parameters (mass, luminosity, effective 
temperature) and abundances of chemical elements, which is comparable 
to classical model atmospheres. 
The effects of stellar pulsation are simulated by time-dependent boundary conditions at 
the inner edge of the com\-pu\-ta\-tion\-al domain, located just below the stellar photosphere. 
The velocity of the innermost mass shell is prescribed as a sinusoidal variation with 
the period and amplitude as parameters (so-called piston model), and the amplitude of the 
accompanying luminosity variation can be adjusted separately 
\citep[see App.\,B in][]{hoefetal16}.

\subsection{Grain-size-dependent opacities}
\label{s:sddo_opac}

\begin{figure}
\centering
\includegraphics[width=\hsize]{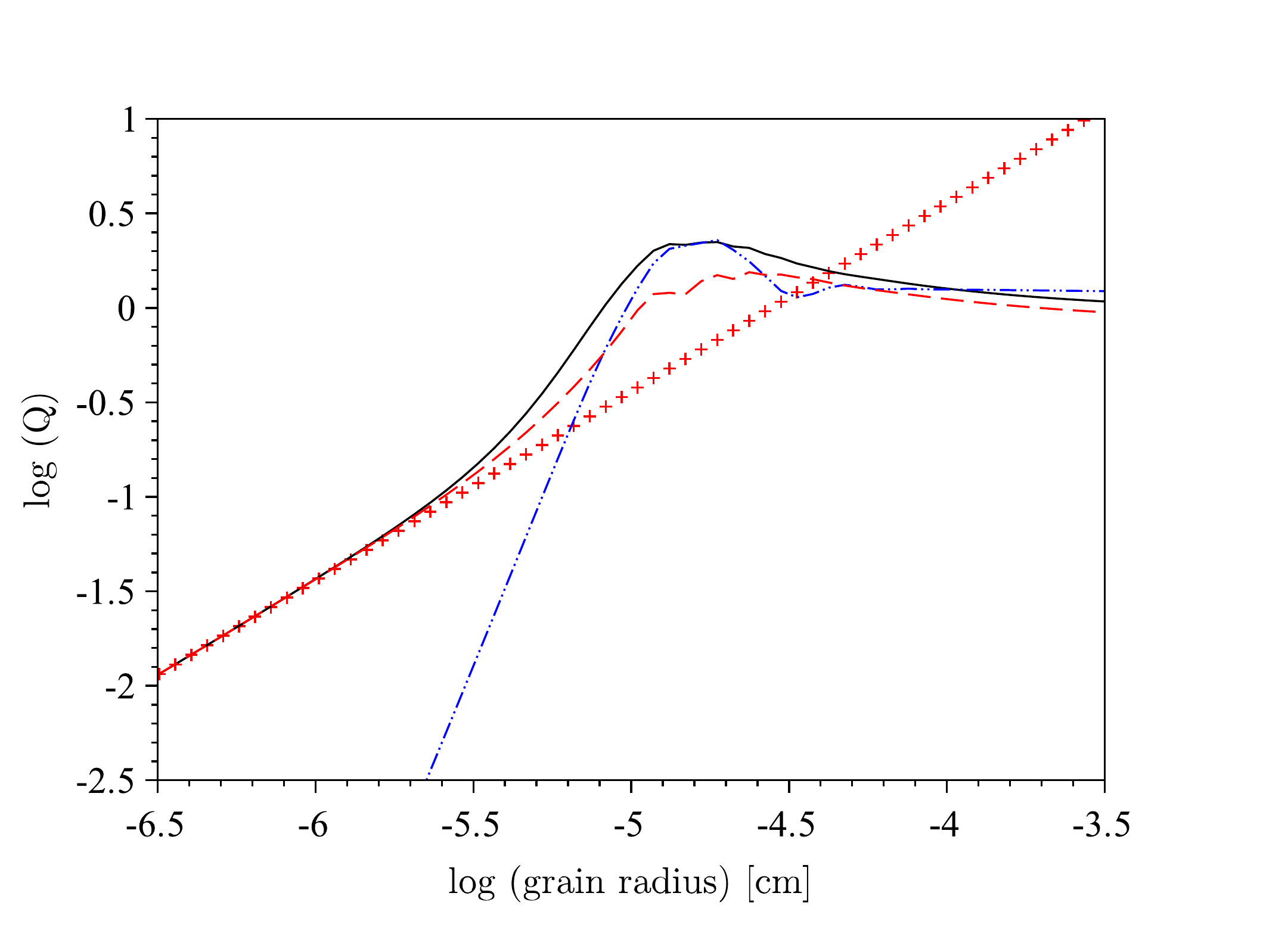}
\includegraphics[width=\hsize]{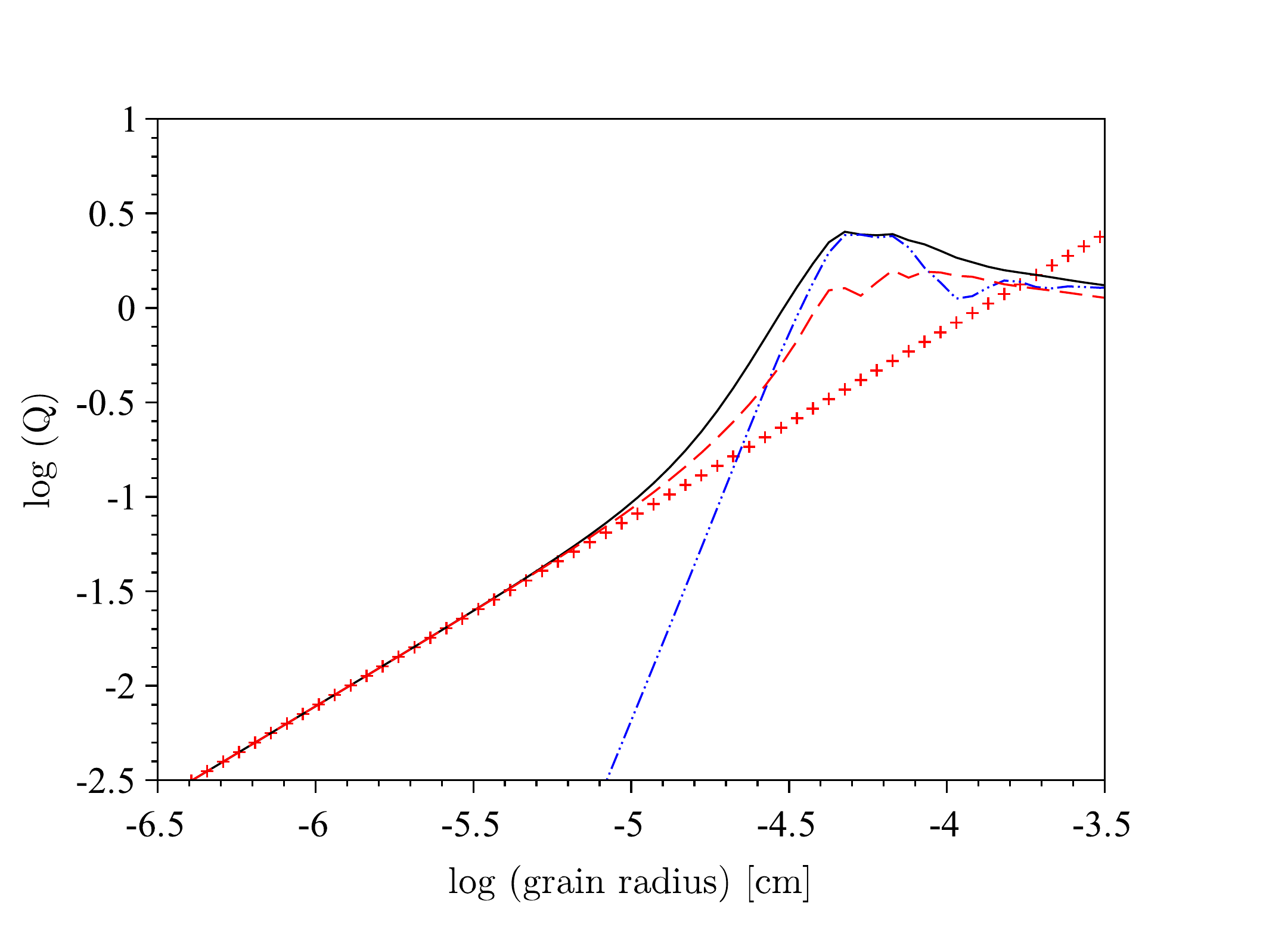}
     \caption{Efficiency factors as a function of grain size for an amC particle computed 
     with full Mie theory (lines, dashed red: absorption, 
     dash-dotted blue: scattering, black: radiative pressure) and in 
     the SPL approximation (absorption, plus signs). 
     Top panel: Values at a wavelength of $0.54 \, \mu$m ($V$ band).
Bottom panel: Values at a wavelength of $2.1 \, \mu$m $(K$ band; 
     see text for details).
     While scattering is negligible in the SPL, it contributes significantly 
     to the radiation pressure for particles with sizes that are comparable to the wavelength.
}

     \label{f_Q_V_agr}
\end{figure}

As mentioned above, the dynamical models presented here differ from those in Paper IV by the 
treatment of dust opacities. 
The total opacity (cross-section/mass) at wavelength $\lambda$ of an ensemble of 
spherical dust grains with radii, $a_{\rm gr}$, embedded in a gas of mass density $\rho$, 
can be expressed as 
\begin{equation}\label{e_kappad}
  \kappa_{\lambda} = \frac{\pi}{\rho} \int_{0}^{\infty} a_{\rm gr}^2 \, Q(a_{\rm gr},\lambda) 
  \, n(a_{\rm gr}) \, da_{\rm gr} ,
\end{equation}
where $n(a_{\rm gr})\,da_{\rm gr}$ is the number density of grains in the size interval 
$da_{\rm gr}$ around $a_{\rm gr}$, and 
$Q(a_{\rm gr},\lambda)$ is the efficiency factor, defined as the radiative cross-section 
$C(a_{\rm gr},\lambda)$ divided by the geometrical cross-section of the grains, that is
\begin{equation}
Q(a_{\rm gr},\lambda) = \frac{C(a_{\rm gr},\lambda)}{\pi \, a_{\rm gr}^2} . 
\end{equation}
In the DARWIN code, the properties of carbon dust, at distance from the centre of $r$ and 
a time of $t$, are described by moments of the grain size distribution, weighted with powers 
of the grain radius:  
\begin{equation}
  K_i \, (r,t) \propto \int_{0}^{\infty} a_{\rm gr}^i \, n(a_{\rm gr}) \, da_{\rm gr} \qquad (i=0,1,2,3),
\end{equation}
where $K_0$ is proportional to the total number density of grains (the integral of the 
size distribution function over all grain sizes), $K_1$ is related to the average 
grain radius, and $K_3$ to the average volume of the grains, respectively.
\footnote{The actual size distribution, $n(a_{\rm gr}) \, da_{\rm gr}$, is not known during 
a simulation run, but can, in principle, be reconstructed afterwards.}
By introducing the quantity $Q' = Q/a_{\rm gr}$, we can re-write Eq.\,(\ref{e_kappad}) as 
\begin{equation}\label{e_kap_qa}
  \kappa_{\lambda} = \frac{\pi}{\rho} \int_{0}^{\infty} Q'(a_{\rm gr},\lambda) 
  \, a_{\rm gr}^3 \, n(a_{\rm gr}) \, da_{\rm gr} ,
\end{equation}
where the integral, apart from the factor $Q'(a_{\rm gr},\lambda)$ looks similar to 
the definition of moment $K_3$. 

For grains that are much smaller than the wavelength under consideration, the small particle 
limit (SPL) of Mie theory gives 
$Q_{\rm abs} \propto a_{\rm gr}$ and $Q_{\rm sca} \propto a_{\rm gr}^4$ 
for the absorption and scattering efficiencies, respectively. 
In this case, $Q'_{\rm abs}$ is independent of the grain radius, and the integral 
over grain size can be represented by moment $K_3$ of the size distribution, while 
the scattering efficiency becomes negligible. 
The assumption of SPL has been used in many dust-driven wind models found in the literature, 
including Papers II -- IV. 
As it turned out, however, the resulting grain sizes may be well beyond the range where 
the SPL is valid. 

For grains that are of sizes comparable to the wavelengths under consideration, the 
quantity $Q'$ depends on the grain radius in a complex way, and can be computed 
numerically from full Mie theory. 
Assuming that the grain sizes at each point in the model are well-represented by the 
average grain radius $\langle a_{\rm gr}\rangle$ 
(derived from moment $K_1$ of the size distribution), we can re-formulate 
Eq.\,(\ref{e_kap_qa}) as 
\begin{equation}\label{e_kap_k1k3}
  \kappa_{\lambda} = \frac{\pi}{\rho} \, Q'(\langle a_{\rm gr} \rangle ,\lambda)  
  \int_{0}^{\infty}  a_{\rm gr}^3 \, n(a_{\rm gr}) \, da_{\rm gr} 
  \propto Q'(\langle a_{\rm gr} \rangle) \, K_3 .
\end{equation}

In the models presented in this paper, the efficiency factors for absorption and 
scattering, as well as the mean scattering angle required for calculating the 
radiative pressure, are computed using Mie theory 
for spherical particles (programme BHMIE by 
B.T.\,Draine\footnote{{\tt www.astro.princeton.edu/\~{ }draine/scattering.html}}) 
and optical properties of amorphous carbon by \cite{RM91}. 
The efficiency factor, $Q_{\rm acc} (a_{\rm gr},\lambda),$ representing radiative pressure 
contains contributions from true absorption and scattering, that is
\begin{equation}
  Q_{\rm acc} = Q_{\rm abs} + (1 - g_{\rm sca}) \, Q_{\rm sca}
,\end{equation}
where $g_{\rm sca}$ is an asymmetry factor describing deviations from isotropic scattering. 
Examples of efficiency factors and their dependence on grain size and wavelength 
are shown in Fig.\,\ref{f_Q_V_agr}, demonstrating the limitations of the SPL assumption, 
used in earlier models.


\subsection{Stellar and pulsation parameters}
\label{s:DMAs}
The models have fundamental parameters (effective temperature, luminosity, and 
stellar mass) as given in Table~\ref{t:gridparam}. 
The stellar parameter combinations in the present grid are the same as in \citet{E+14}.
All models have solar abundances following \cite{AsGS05}, except for the carbon abundance.
The values are given on the scale where log N$_H$ $\equiv$  12.00. 
The abundances from \cite{AsGS05} correspond to a composition by mass: 
X/Y/Z = 0.73/0.25/0.015--0.020; the variation being due to the varying carbon abundance.
For every combination of stellar parameters in the table, models were computed with 
carbon excesses of log(C--O) + 12 = 8.2, 8.5 and 8.8
\footnote{
The relation between the carbon excess, log(C--O)+12, and the commonly
used quantity C/O is given in Table~\ref{t:coratios}. 
In contrast to other papers in the literature, we used the carbon excess to characterise 
the models, because this quantity directly translates into the amount of carbon available 
for the formation of carbon-bearing molecules (other than CO) and dust grains.
}
and with piston velocity amplitudes of $\Delta u_{\rm p}$ = 2, 4 and 6 km\,s$^{-1}$.
The pulsation periods are given by the luminosity through a period-luminosity relation following
\cite{FeGWC89}; for easy reference, the periods are also given in Table~\ref{t:gridparam},
although they are not treated as an independent parameter. 
\footnote{
The $P-L$ relation by \cite{FeGWC89} is based on Miras in the LMC. 
In a diagram of $K$ magnitudes versus period as, for example, in \cite{Ita04}, our models would 
largely overlap with the observed stars in sequence C. The stars belonging to this sequence 
are commonly believed to be fundamental mode pulsators \citep{Wood99}.
}

\begin{table}
\begin{center}
\caption{Combinations of fundamental stellar parameters covered by the model grid. 
We note that the period is not an independent parameter, but it is coupled to the stellar luminosity (see text).
For each set of parameters listed here, we varied the velocity amplitude
($\Delta u_{\rm p}$) at the inner boundary.}
\begin{tabular}{cccll}
\hline
\hline
$T_\star$   & log $L_\star$  & $P$ & $M_\star$     & log(C--O)+12  \bigstrut[t] \\
 {[K]}          &  [$L_\odot$]    & [d]    & [$M_\odot$]  &  [dex] \bigstrut[b] \\
\hline
2600 & 3.70  & 294 &  0.75, 1.0    & 8.2, 8.5, 8.8  \bigstrut[t] \\
&         3.85  &  390 & 0.75, 1.0, 1.5, 2.0  & 8.2, 8.5, 8.8  \\
&         4.00  &  525 & 1.0, 1.5, 2.0    & 8.2, 8.5, 8.8  \bigstrut[b] \\
\hline
2800 & 3.55  & 221 & 0.75    & 8.2, 8.5, 8.8  \bigstrut[t] \\
&         3.70  &  294 & 0.75, 1.0    & 8.2, 8.5, 8.8  \\
&         3.85  &  390 & 0.75, 1.0, 1.5, 2.0   & 8.2, 8.5, 8.8  \\
&         4.00  &  525 & 1.0, 1.5, 2.0   & 8.2, 8.5, 8.8  \bigstrut[b] \\
\hline
3000 & 3.55  &  221 & 0.75    & 8.2, 8.5, 8.8  \bigstrut[t] \\
&         3.70  &  294 & 0.75, 1.0   & 8.2, 8.5, 8.8  \\
&         3.85  &  390 & 0.75, 1.0, 1.5, 2.0   & 8.2, 8.5, 8.8 \\
&         4.00  &  525 & 1.5    & 8.2, 8.5, 8.8   \bigstrut[b] \\
\hline
3200 & 3.55  & 221 &  0.75   & 8.2, 8.5, 8.8  \bigstrut[t] \\
&         3.70  & 294 &  0.75, 1.0    & 8.2, 8.5, 8.8  \bigstrut[b]  \\
\hline
\end{tabular}
\label{t:gridparam}
\end{center}
\end{table}

During a pulsation cycle, the radius of the innermost mass shell of the model and the 
luminosity at that position vary simultaneously, simulating a coupling between the 
variable brightness and size of the star. 
Without a quantitative model of the pulsating stellar interior, assumptions have to be 
made about the forms, the amplitudes, and the relative phases of these temporal variations. 
Regarding the dynamic boundary (i.e. gas velocity), the functional form is of minor importance 
since the outward-travelling waves quickly develop into shocks and the kinetic energy 
transferred into the atmosphere is mostly determined by the velocity amplitude. 
Therefore, we adopted the common assumption of a sinusoidal variation, parametrised by the 
velocity amplitude, $\Delta u_{\rm p,}$ and the pulsation period, $P$. 
In this differential study of the influence of size-dependent dust opacity, we keep the 
choice of \cite{E+14} and assume that the luminosity at the position of the innermost 
mass shell varies in phase with its radius and that the amplitude of the luminosity variation 
is correlated with the radius amplitude. 
An adjustable factor, $f_{\rm L,}$ was introduced \citep[see][for details]{NowHA10} to scale 
the luminosity variations. 
\cite{E+14} varied the value of this parameter and found no significant changes in 
the dynamic properties, so we kept their choice of $f_{\rm L}$~=~2.

\begin{table}
\begin{center}
\caption{Relation between the carbon excess measure log(C--O)+12 and the C/O ratio for the
adopted chemical composition \citep{AsGS05}, i.e. using an oxygen abundance of 
log(N$_O$/N$_H$)+12 = 8.66.}
\begin{tabular}{ccc}
\hline\hline
log(C--O)+12 & C/O & label \bigstrut[t]\bigstrut[b] \\
\hline
8.2 & 1.35 & C5 \bigstrut[t]\\
8.5 & 1.69 & C6 \\
8.8 & 2.38 & C7  \bigstrut[b] \\
\hline
\end{tabular}
\label{t:coratios}
\end{center}
\end{table}

\subsection{Synthetic spectra and photometry}
\label{s:synthspecphot}

For each combination of stellar and pulsational parameters, snapshots of the atmospheric 
structures at various phases (typically 20 per period) during several consecutive periods 
(typically four) and for a few different epochs during the simulation time span were
used for the detailed {\it \emph{a~posteriori}} radiative transfer calculations. 
We need more than one period to obtain a representative behaviour since the 
models are not strictly periodic, mainly due to the time-dependent dust formation.
We applied the code {\tt COMA}
\citep[][and Paper~I]{Aring00}, version 11, to compute atomic, molecular, and amorphous 
carbon (amC) dust. 
While \cite{E+14} used dust opacities calculated under the assumption of the small particle 
limit, here we used size-dependent opacities based on Mie theory and amC optical 
properties from \cite{RM91}.
We note that the treatment of the opacities is consistent with that in the generation
of the dynamic models, except for a higher spectral resolution in the {\it \emph{a~posteriori}}
radiative transfer calculations (10\,000).
 
The resulting synthetic spectra in the 0.35\,--\,25~$\mu$m wavelength range were used to 
compute filter magnitudes in the Johnson-Cousins $BVRI$ system \citep{Besse90} 
and the Johnson-Glass $JHKLL'M$ system \citep{BB88}. 
More details can be found in previous articles, for example Paper~II.
The results given in the table in Appendix~\ref{a:overviewdata}
include a temporal mean of the $V$ and $K$ magnitudes, as well as
their (maximum) ranges, and also the mean values of the colours 
\mbox{($V$\,--\,$I$)}, \mbox{($V$\,--\,$K$)}, \mbox{($J$\,--\,$H$),} and \mbox{($H$\,--\,$K$)}.

 \begin{figure}[h!]
  \resizebox{1.39\hsize}{!}{\includegraphics{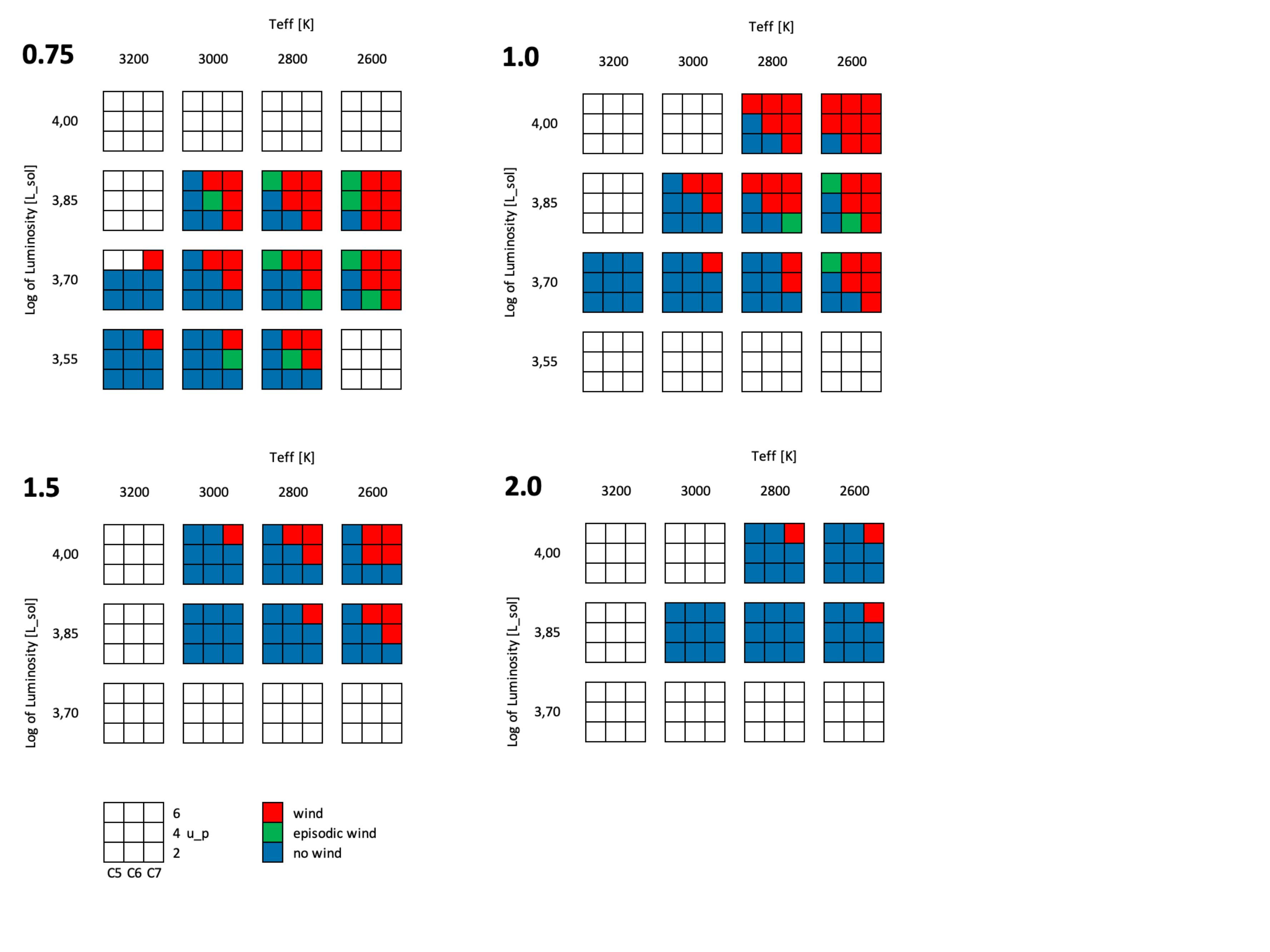}}
  \resizebox{1.39\hsize}{!}{\includegraphics{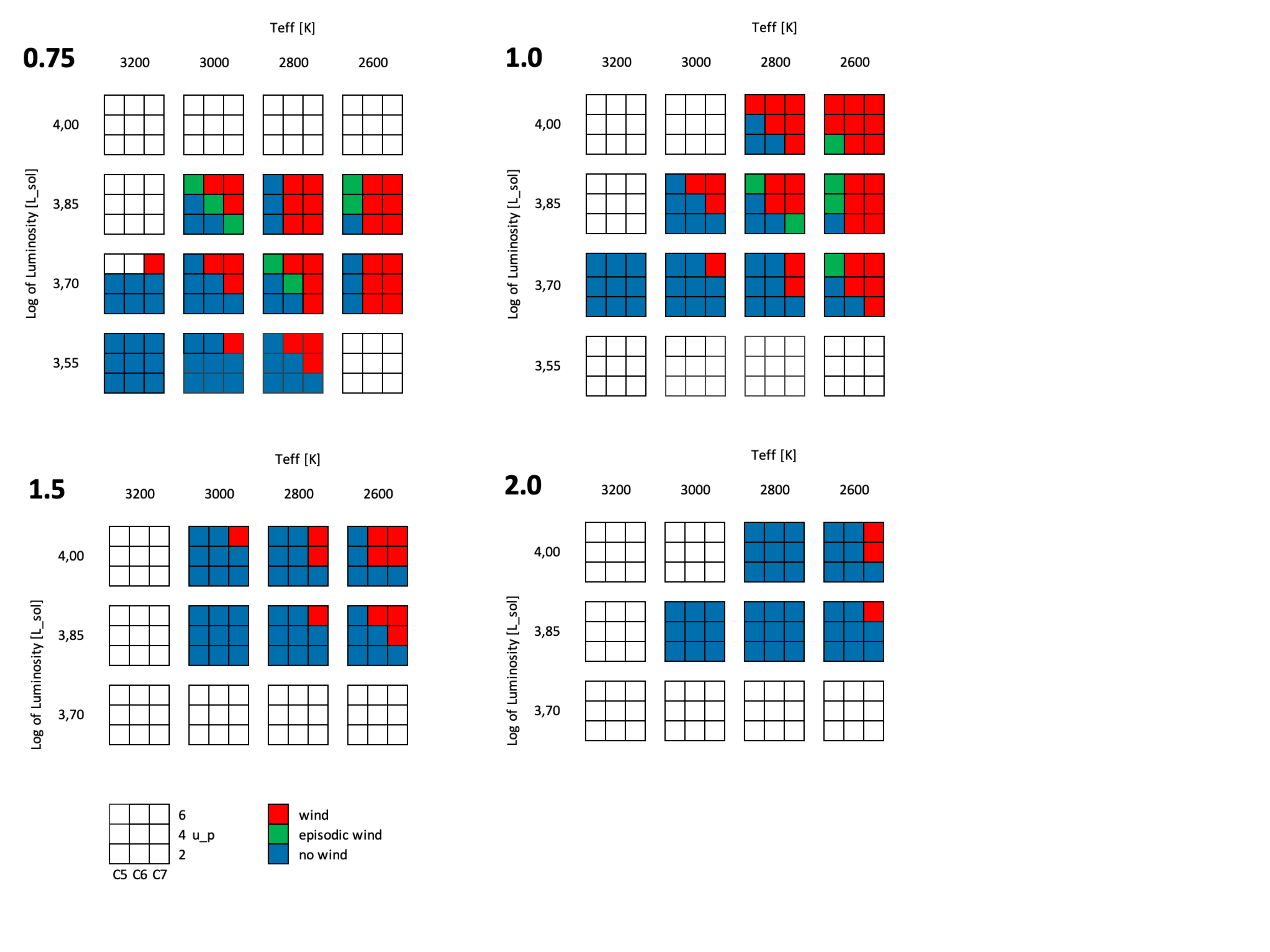}}
  \caption{Dynamic behaviour of models these are also referred to as wind-maps. 
  In an HR-type display, the models are colour-coded as red for wind, 
  blue for no wind, and green for episodic behaviour.
  The upper four panels show the SDDO case and the lower four show
  the grid with the SPL approximation.
  The four sub-panels for each grid show models of different stellar masses 
  indicated in the top left corner. 
  For each given combination of temperature and luminosity, the
  different carbon excesses and piston amplitudes are arranged as shown in the bottom legend.
  }
  \label{windmaps}
\end{figure}


\section{Effects of size-dependent dust opacities}
\label{s:DMAdyn}

In this section, we describe the resulting differences between models using size-dependent 
dust opacities (SDDO) and the SPL assumption.
We mainly discuss the changes of time-averaged properties, giving an overview of trends for the 
whole grid. 
In some cases, we show the time-dependent behaviour of individual models in order to
demonstrate the underlying physical mechanisms.

\subsection{Dynamical and dust properties}
\label{s:DMAdyndust}

The resulting dynamic properties of the new SDDO models are compiled in
the table in Appendix~\ref{a:overviewdata}.
The listed values are time-averages of the mass-loss rates, outflow velocities, 
and carbon condensation degrees, and these were determined in the following way. 
For a given model, the existence of a wind was defined as the outer boundary having reached 
25 stellar radii and the velocity of the gas at the outer boundary being positive (i.e. an outflow).
Then, for each (pulsation) cycle with the wind condition fulfilled, the mass-loss rate and so on were 
determined, and later means and standard deviations were calculated. 
For most models, several hundred cycles were used to determine the means.
For some  combinations of fundamental parameters, a wind is sustained only for limited periods,
resulting in episodic behaviour. 
In this paper, we define a model as episodic when wind conditions prevail for more than 15\% but
less than 85\% of the total time interval (after the first instance of wind conditions).

In general, mass loss is favoured by low $T_\star$ (easier to form dust at cooler temperatures),
high luminosity (more momentum transferred from radiation to dust grains), small stellar mass
(shallower potential well), large (C--O) (more free carbon to form amorphous carbon grains), and 
a large piston velocity amplitude (stellar layers reach out to a greater distance from the 
centre of the star during pulsations).
In total, 268  different dynamic model atmospheres were computed (96 of them produced 
winds and 172 did not lead to outflows). 
For the SPL case, the corresponding numbers are 95 with outflows and 173 without.
An overview of the wind or no-wind status of the different models can be given in so-called 
wind maps, where the status of a model (no-wind/episodic/wind) is shown  
in an HR-diagram-like representation.
Wind maps of the current grid (SDDO, with size-dependent dust opacities) and the SPL grid 
are displayed in Fig.~\ref{windmaps}. 
As seen, the qualitative changes in dynamic behaviour are minor, mostly consisting of 
changes from/to the episodic cases.

\begin{figure}
  \resizebox{\hsize}{!}{\includegraphics{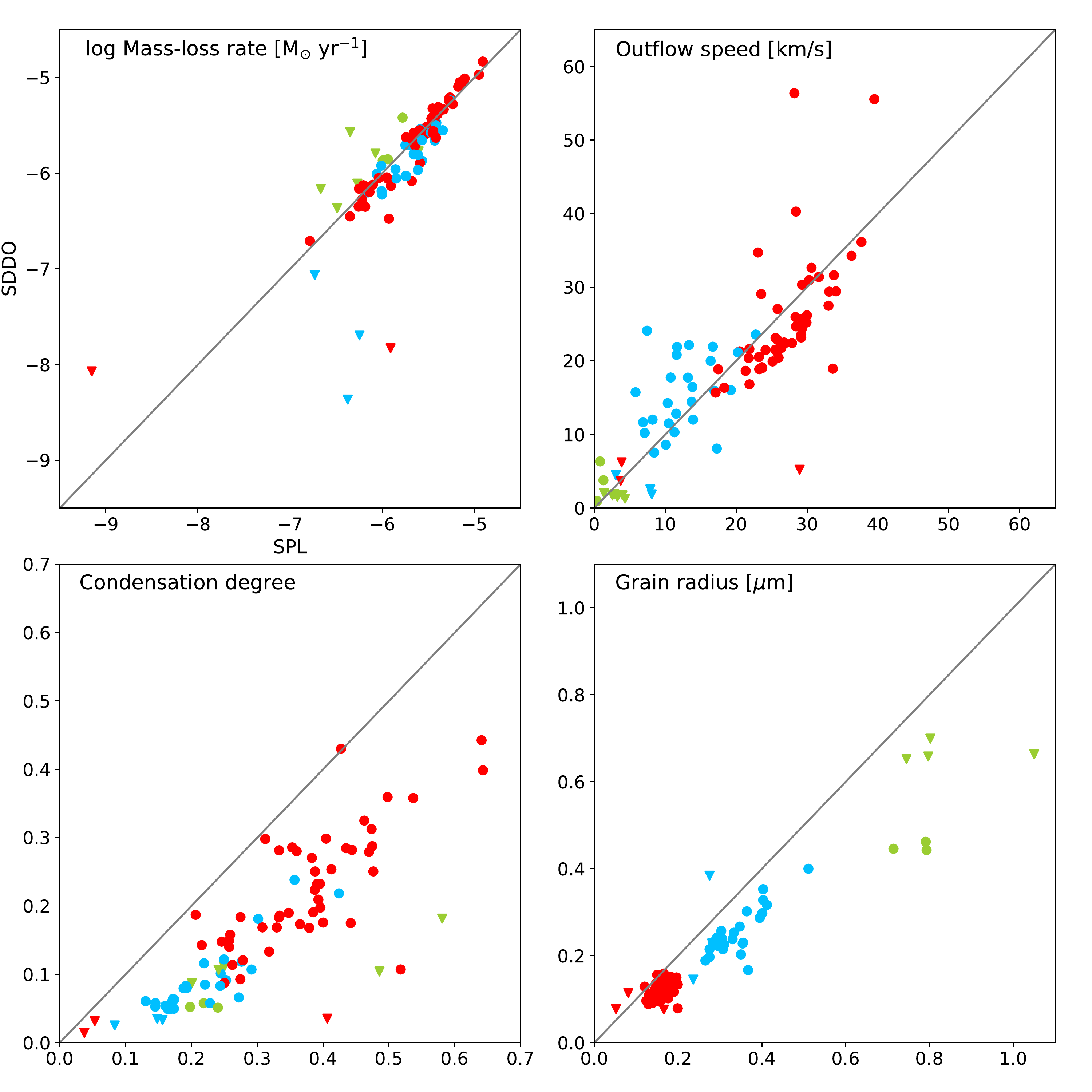}}   
  \caption{Comparison of different properties of SDDO versus SPL grids.
  The quantities plotted are mass-loss rates (upper, left), outflow speeds (upper, right), 
  condensation degrees (lower, left) and dust grain radii (lower, right). 
  The different carbon excesses are shown in colour:  log(C-O)=8.2 in green, 
  8.5 in blue, and 8.8 in red. Circles denote wind models and triangles denote episodic models.
  \label{figDMMa1s}}
\end{figure}

The relatively small effects on the mass-loss rates and on the outflow velocities when we 
replace SPL by SDDO are clearly seen in the upper two panels of 
Fig.~\ref{figDMMa1s}, where the new (SDDO) values are plotted against the SPL ones. 
For some model parameter combinations and borderline cases, the wind turns on and off 
one or more times during the simulation sequence(s); 
these episodic cases are plotted as triangles where one or both of the models display this behaviour.
As seen, it is only these episodic cases that exhibit large differences in 
mass-loss rates between SPL and SDDO.
We note the expected trend of higher outflow speeds for the 
larger values of the carbon excess (more dust giving more acceleration to the flow).

With SDDOs, the carbon condensation degrees and grain sizes  are smaller than for the SPL models. 
This is seen in Fig.~\ref{figDMMa1s} for all non-episodic models producing winds.
The condensation degree of carbonaceous dust, and hence the dust-to-gas ratio, is roughly half 
as big in the SPL case.
We note that the dust grains do not grow so large as they quickly
travel through the dust formation regions, which is a consequence of higher radiative force
compared to the SPL case (see Fig.~\ref{f_Q_V_agr}).
The largest grains grow in the slow outflows in models with small carbon excesses, 
especially for the episodic cases.

\begin{figure}
  \resizebox{\hsize}{!}{\includegraphics{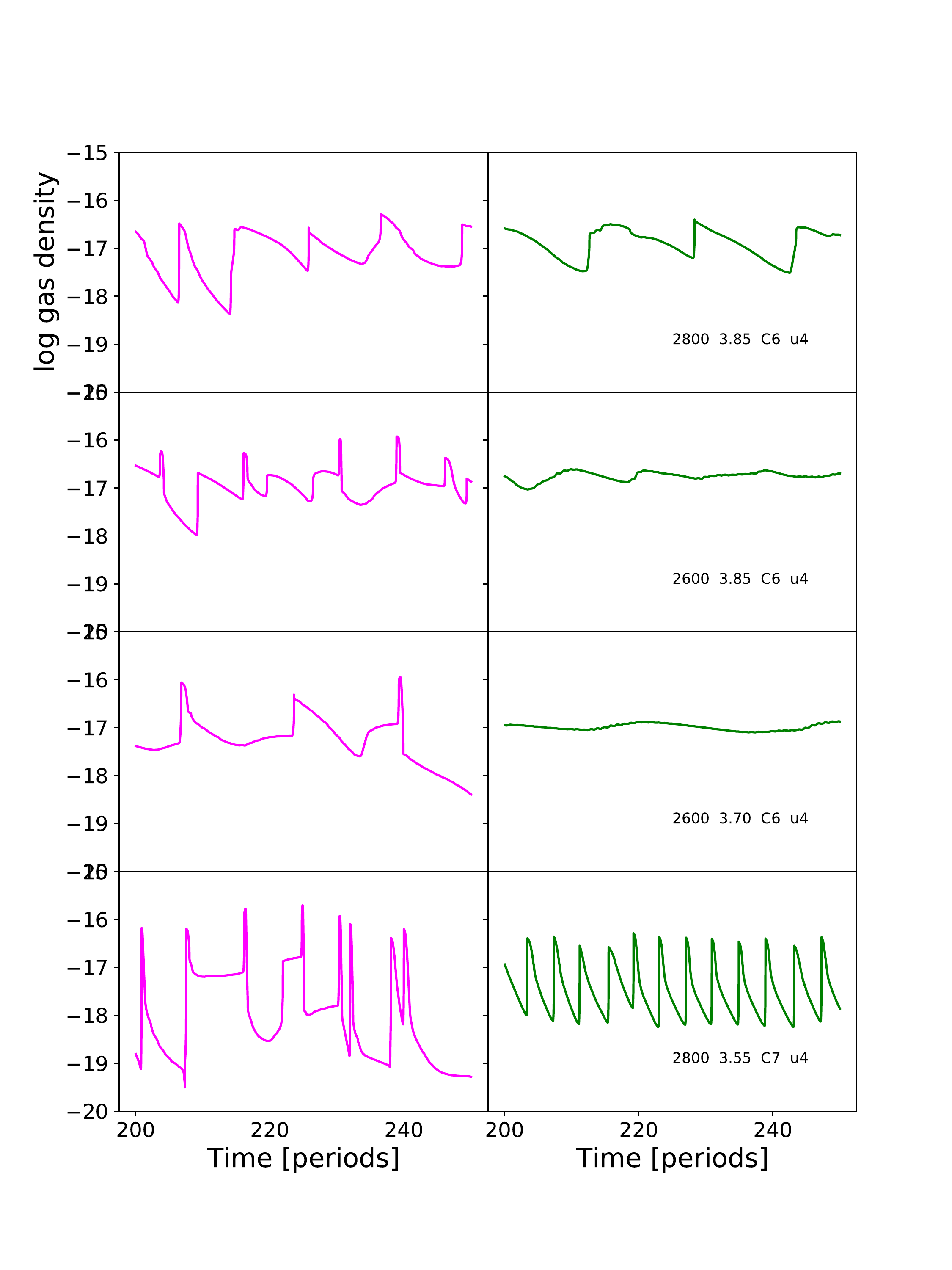}} 
  \caption{Gas density at outer boundary as a function of time for selected 
  models with a stellar mass of 0.75 M$_{\odot}$. The left column displays the SDDO values, 
  the right one the SPL values. 
  Stellar parameters, T$_{\rm{eff}}$, log L, carbon excess class and
  piston amplitude, are given in the SPL panels. 
  The densities are given in cgs units: T$_{\rm{eff}}$ in K, 
  log L in L$_\odot,$ and piston amplitudes in km\,s$^{-1}$.
  We use a fixed range in gas density in all panels. 
  }
  \label{figDMM_rhoext_m075}
\end{figure}

\begin{figure}
  \resizebox{\hsize}{!}{\includegraphics{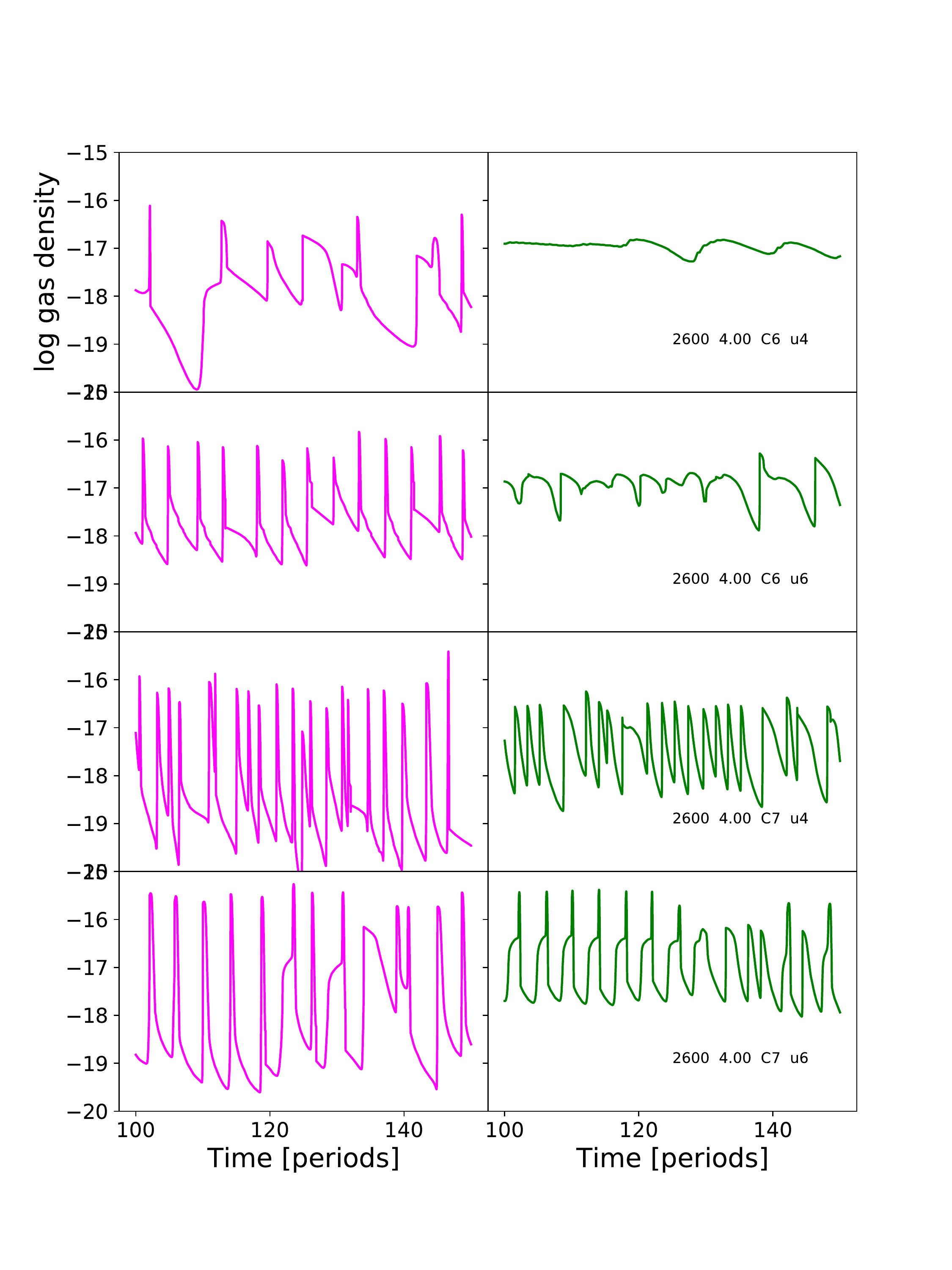}}    
  \caption{Gas density at the outer boundary as a function of time for 1.5 M$_{\odot}$
  models (see Fig.~\ref{figDMM_rhoext_m075} for details).}
  \label{figDMM_rhoext_m15}
\end{figure}

The change in dust opacity from SPL to SDDO causes changes in the time-dependent behaviour 
of the models.
This is visualised in Fig.~\ref{figDMM_rhoext_m075} for some models with stellar 
masses of 0.75 M$_{\odot}$ and for some models with masses of 1.5 M$_{\odot}$ 
in Fig.~\ref{figDMM_rhoext_m15}, where the (gas) density
at the outer boundary for 50 consecutive periods are displayed.
We see the same general behaviour for all cases:  a larger density contrast in the 
wind for the SDDO models as compared to the SPL ones, that is a wind with narrower and 
more pronounced gusts of gas and dust.
This is probably a consequence of the steeper dependence of radiative acceleration on the size 
of the growing grains in the SDDO case (see Figure~\ref{f_Q_V_agr}).

In summary, the mass-loss rates and outflow speeds are not strongly
affected by replacing the SPL approximation with SDDOs. The higher radiative force on sub-micron-sized dust grains, 
compared to the SPL case, however, tends to result in smaller dust grains and lower
condensation degrees. These general trends in wind properties are in good agreement with the
results of a pilot study by \citet{MatH11}, which is based on a much smaller number of DARWIN models
than in the current grid.


\subsection{Photometric properties}
\label{s:DMAphot}

\begin{figure}
  \resizebox{\hsize}{!}{\includegraphics{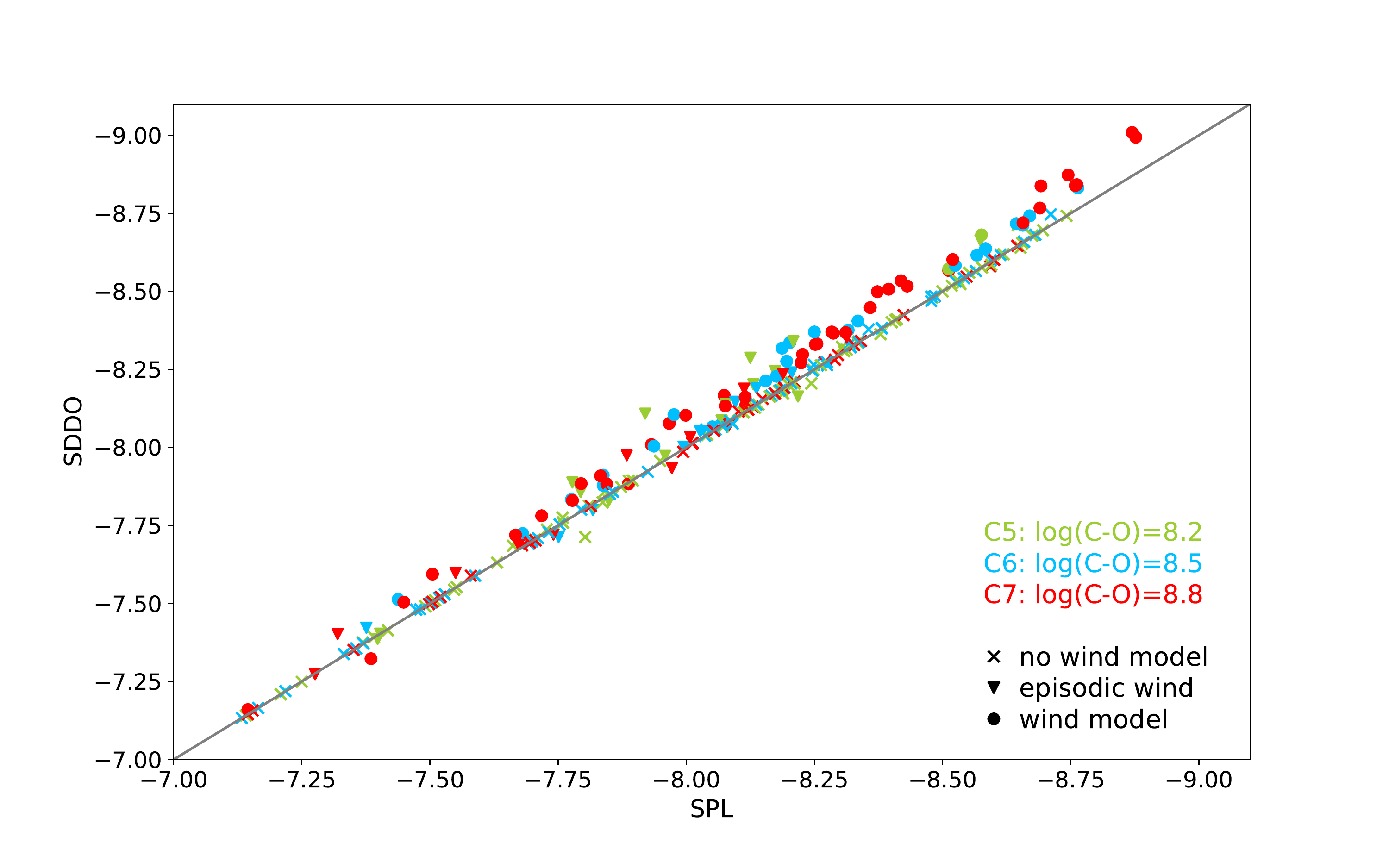}}   
  \caption{Mean $K$ magnitude for models in the entire grid computed without
  using the dust opacity in the {\it \emph{a~posteriori}} radiative transfer (COMA). 
  The values for the SDDO cases versus the SPL ones are plotted.}
  \label{figDMMa1asKmagndus}
\end{figure}

\begin{figure}
  \resizebox{1.05\hsize}{!}{\includegraphics{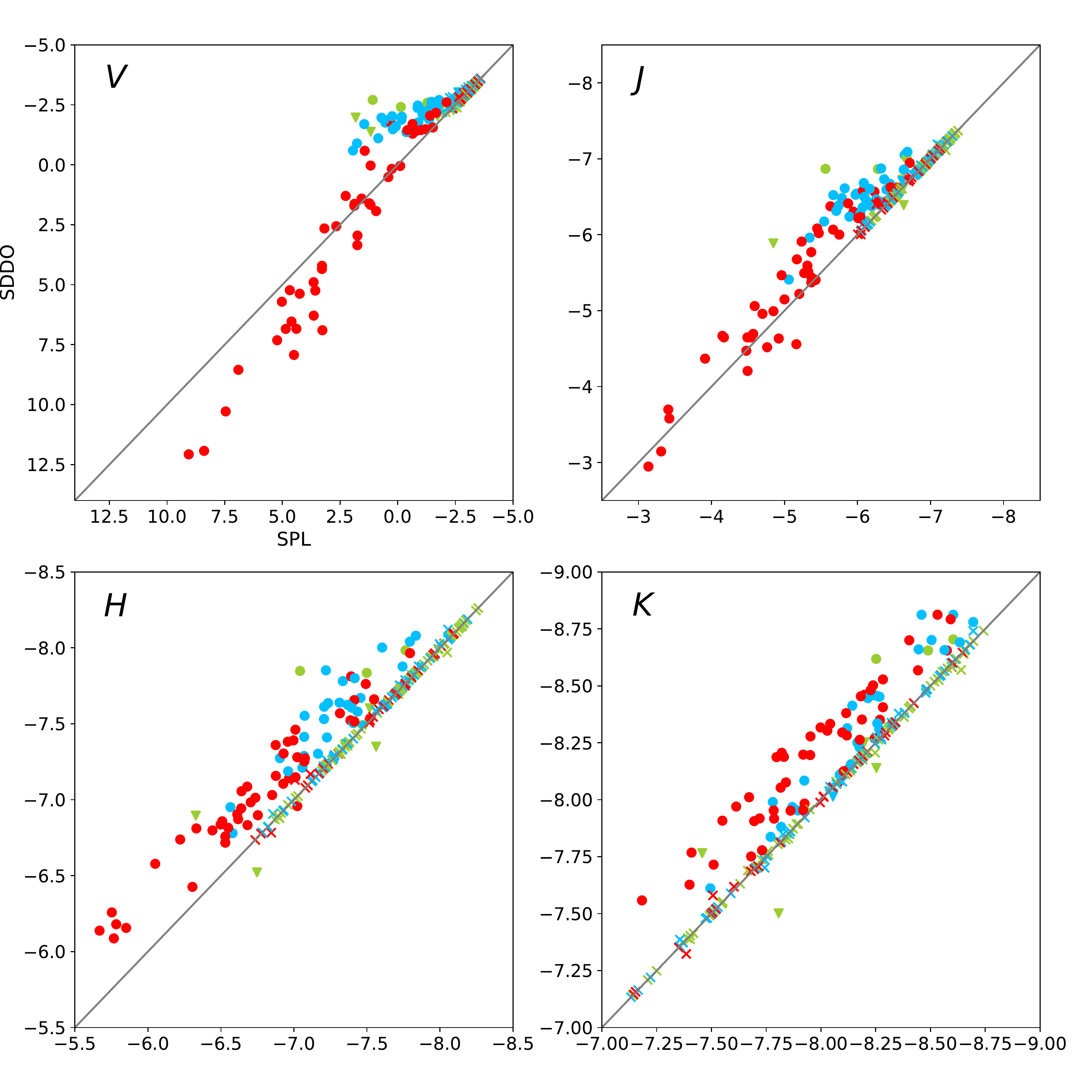}}   
  \caption{Mean $V$, $J$, $H,$ and $K$ magnitude for models in the entire grid. 
  The values for the SDDO cases versus the SPL ones are plotted.
  Colours and symbols are explained in Fig.~\ref{figDMMa1asKmagndus}.}
  \label{figDMMa1as_mag}
\end{figure}
\begin{figure}
  \resizebox{1.15\hsize}{!}{\includegraphics{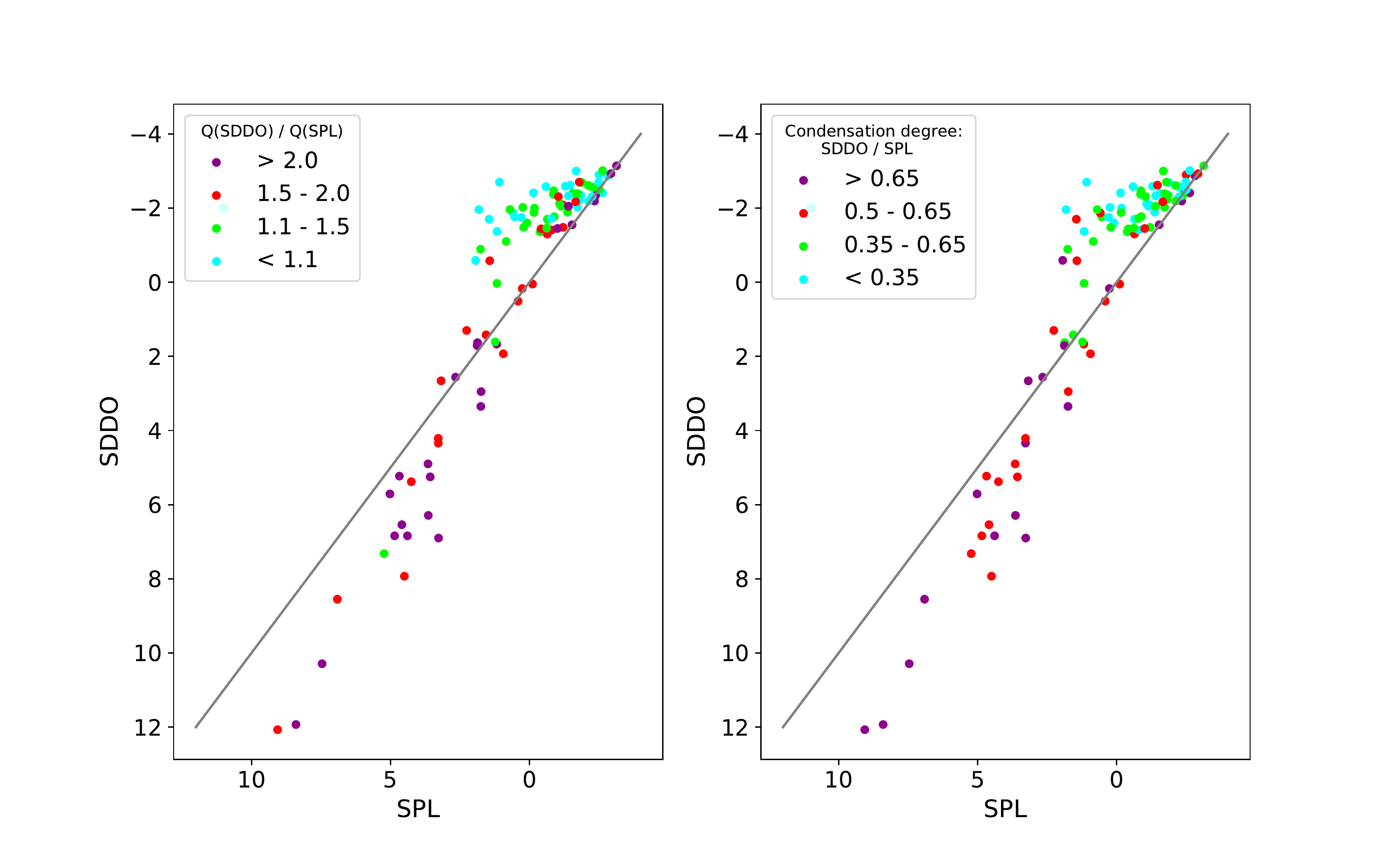}}   
  \caption{Mean $V$ magnitude for wind models in the SDDO versus SPL cases. 
  In the left panel, the points are colour-coded
  according to Q(SDDO)/Q(SPL) and evaluated at appropriate grain radii.
  In the right panel, the colour is given by the ratio of the mean condensation degree in 
  the SDDO and the SPL cases.}
  \label{figDMM_V_Qfact_fcfact}
\end{figure}

Going from SPL to SDDO, the photometric magnitudes could change due to the direct effects of 
the different dust opacities or through the indirect effects on the structures of 
the atmosphere+wind caused by the dust.
The effects due to changes in the structures can be estimated by computing spectra without taking
opacity from the dust into account. We find that the impacts of structural changes 
are minor for all filters
($\lesssim$ 0.1 mag);
Fig.~\ref{figDMMa1asKmagndus} shows a comparison of the $K$ magnitudes.

An overview of the differences in the photometric properties with SDDOs compared to SPL models
is shown in Fig.~\ref{figDMMa1as_mag}, displaying the mean $V$, $J$, $H$,
and $K$ magnitudes for the models in the entire grid (wind and no-wind models).
The $K$ magnitude is typically brighter by 0.2 -- 0.4  magnitudes for the wind models with SDDO.
This is mainly due to the reduced amount of dust (lower condensation degree) in the SDDO models
(see Fig.~\ref{figDMMa1s}). 
The $H$ magnitudes for the wind models
are also generally brighter without SPL, typically by 0.5 magnitudes.
From the $J$ magnitude to shorter wavelengths,  the influence of
dust is more complex; for small dust veiling the value is
smaller (i.e. brighter), but for increasing obscuration the magnitudes are
fainter for the SDDO models, in $V$ by up to three magnitudes.
As seen in the figure, this occurs primarily for the highest carbon excess.

For the $V$ band, we estimated the effect by comparing the Q value for absorption 
(Fig.~\ref{f_Q_V_agr}, upper panel) for the smaller grains 
in the SDDO case with the Q value for the bigger grains in SPL models 
(see Fig.~\ref{figDMMa1s}). 
We find that the former is almost always larger.
However, for some models (mostly with a small carbon excess, C5) where the
mean grain radius is larger than 0.4 $\mu$m (log a$_{\rm gr}$ = -4.4 cm) in the SPL case, 
we see in Fig.~\ref{f_Q_V_agr} (upper panel)
that Q(SDDO) is smaller regardless of the SDDO grain radius.
In Fig.~\ref{figDMM_V_Qfact_fcfact}, we again plot $V$(SDDO) versus $V$(SPL), now colour-coded 
by the value of Q(SDDO)/Q(SPL) calculated at the appropriate mean grain radii.
The cases with the lowest SDDO/SPL ratios of absorption efficiency tend to
show brighter mean $V$ magnitudes.
The difference in $V$ is also dependent on the carbon condensation degree. 
This is shown in the right-hand panel of Fig.~\ref{figDMM_V_Qfact_fcfact}, 
where $V$ is colour-coded according 
to the ratio of the mean condensation degree in the SDDO model to that in the SPL model.
As expected, a lower relative amount of dust correlates well with brighter mean $V$
magnitudes.
In summary, the behaviour of the $V$ magnitude in Fig.~\ref{figDMMa1as_mag} is the result 
of the combined effects of changed Q values and different condensation degrees,
as illustrated in Fig.~\ref{figDMM_V_Qfact_fcfact}.

\begin{figure}
  \resizebox{1.1\hsize}{!}{\includegraphics{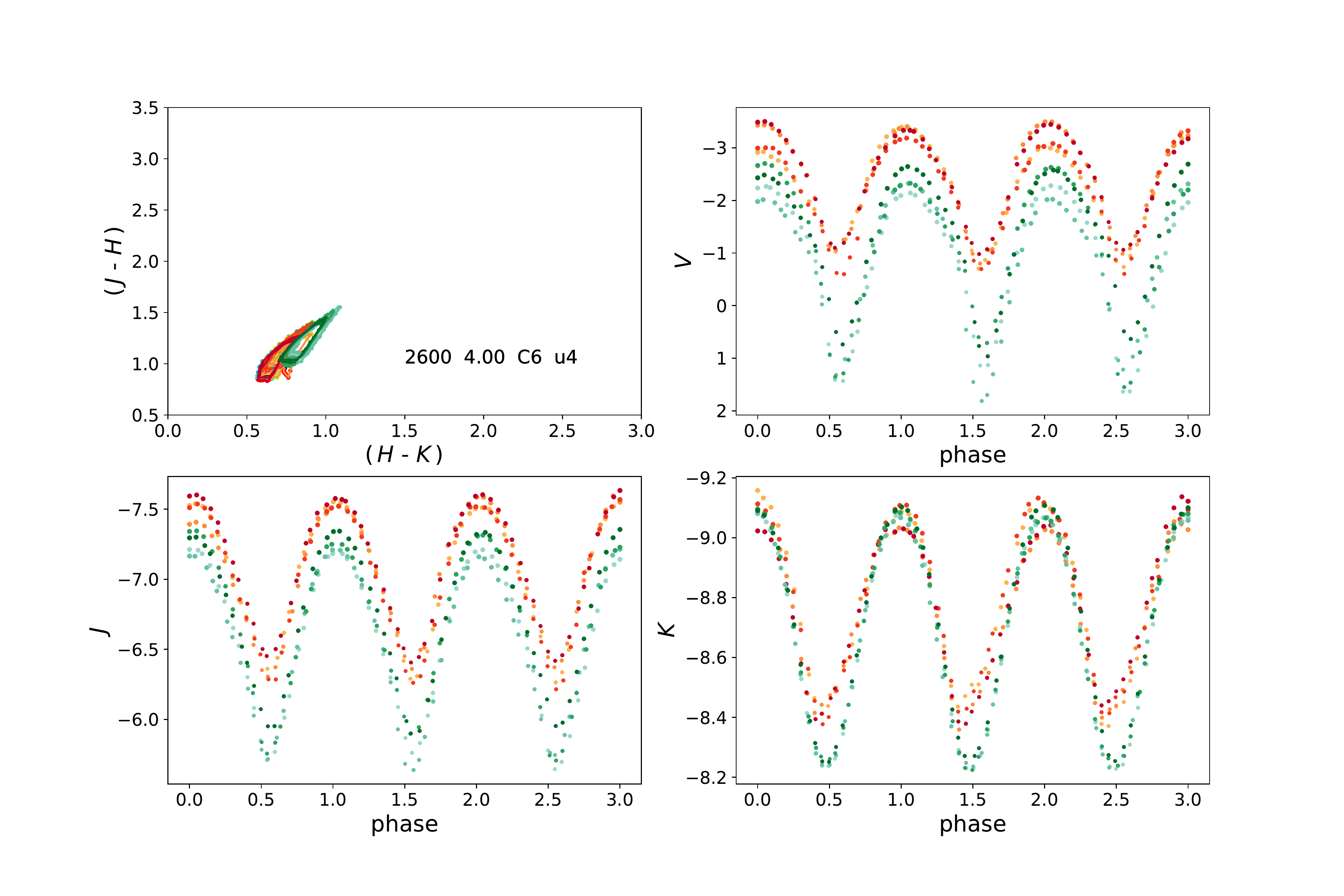}}   
  \caption{Photometric variations of the models shown in the top row of 
  Fig.~\ref{figDMM_rhoext_m15}.
  The \mbox{($J$\,--\,$H$)} versus \mbox{($H$\,--\,$K$)} diagram (top, left) 
  and the light curves  
  of the $V$ (top, right), $J$ (bottom, left), and $K$ (bottom, right) 
  magnitudes for three periods in several epochs. 
  The SDDO values are plotted in orange--red hues, and the SPL values are plotted in blue--  ones. 
  The model has the parameters T$_{\rm eff}$ = 2600$\,$K, log L = 4.0\,L$_\odot$, 
  mass 1.5 M$_\odot$, log(C-O) = 8.5, and a piston
  amplitude of 4 km$\,$s$^{-1}$.}
  \label{figDMM_lc2c_64}
\end{figure}
\begin{figure}
  \resizebox{1.1\hsize}{!}{\includegraphics{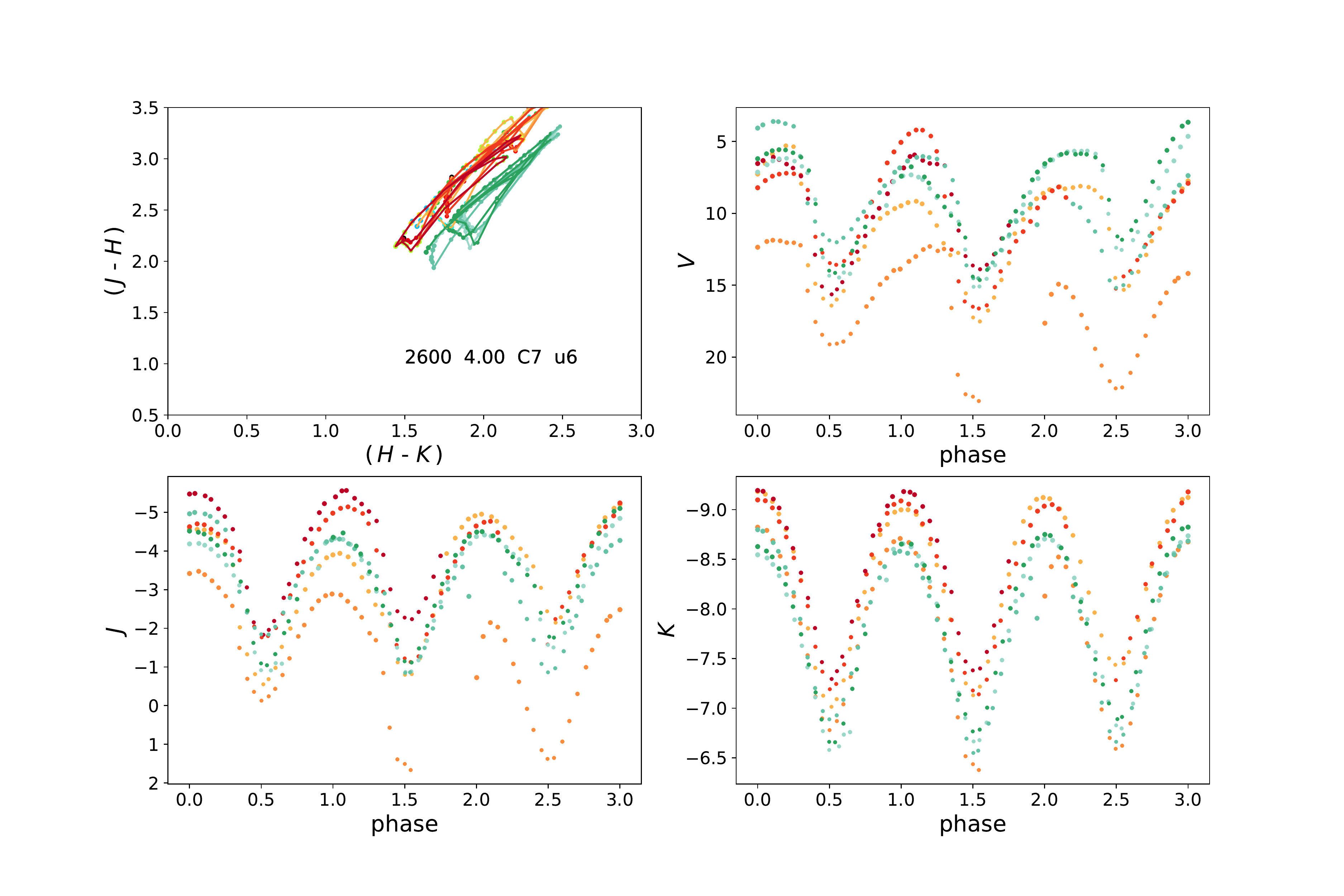}}   
  \caption{As Fig.~\ref{figDMM_lc2c_64}, but for the models with log(C-O) = 8.8 and piston
  amplitude of 6 km$\,$s$^{-1}$ (bottom row in Fig.~\ref{figDMM_rhoext_m15}).}
  \label{figDMM_lc2c_76}
\end{figure}

Turning from the mean magnitudes to the time-variations, that is, the light curves,
we show some examples in Fig.~\ref{figDMM_lc2c_64} and Fig.~\ref{figDMM_lc2c_76}.
To illustrate the photometric variations, we chose two models in
Fig.~\ref{figDMM_rhoext_m15}
with different characteristics: one with a smaller carbon excess and pulsation amplitude, 
the other with larger values for these properties; otherwise, the models have the same 
fundamental parameters.
In Fig.~\ref{figDMM_lc2c_64} and Fig.~\ref{figDMM_lc2c_76}, we show the light-curves 
in $V$, $J,$ and $K$ as well as the two-colour diagram for 
\mbox{($J$\,--\,$H$)} versus \mbox{($H$\,--\,$K$)}  for the two models, for both SDDO and SPL. 
We note period-to-period variations that are larger at visual compared to 
near-infrared (NIR) wavelengths.
The same is true for epoch-to-epoch variations. 
For high-mass-loss models (such as in Fig.~\ref{figDMM_lc2c_76}),
the influence of dust may lead to less periodic variations in $V$.

\begin{figure}
  \resizebox{1.05\hsize}{!}{\includegraphics{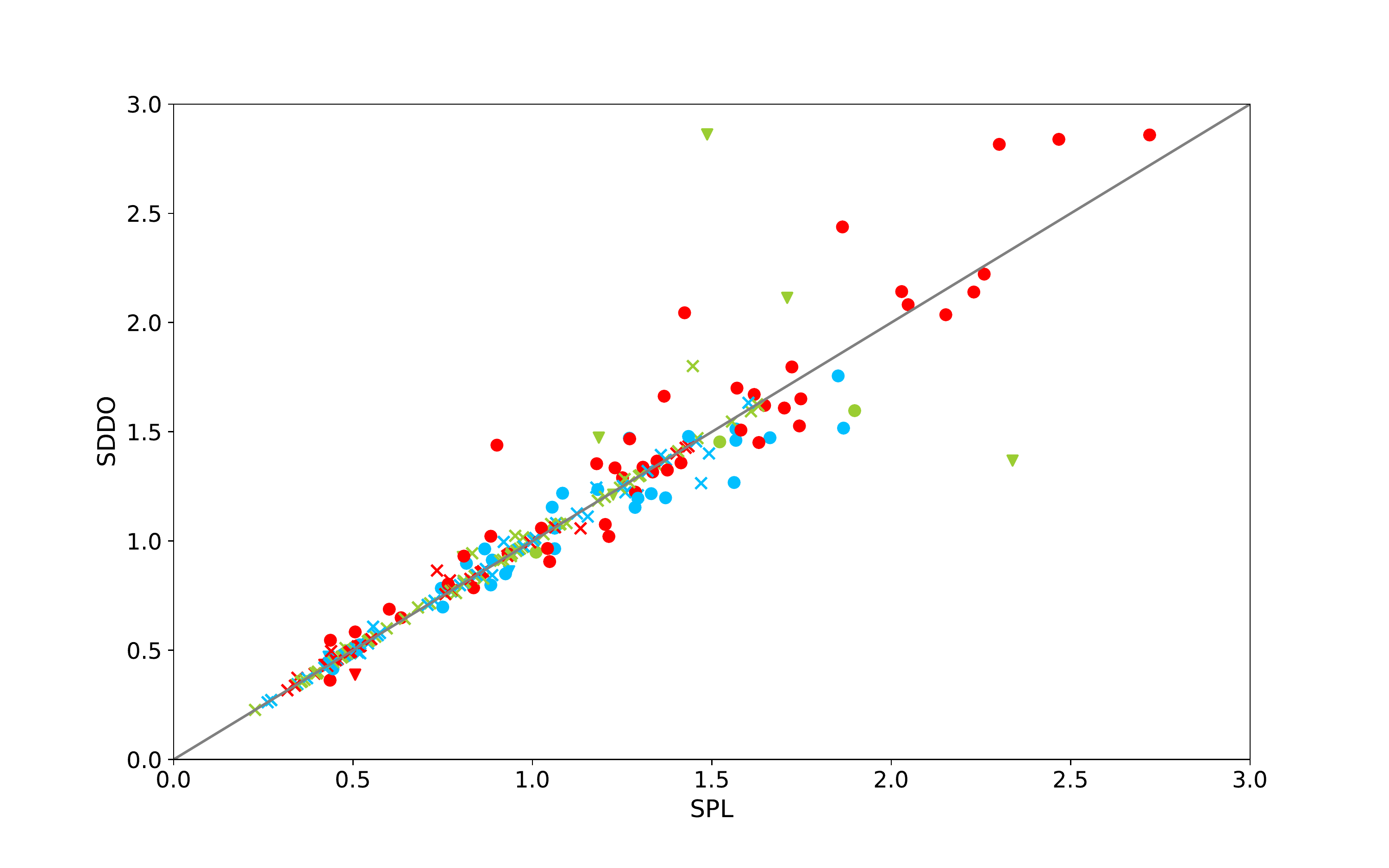}}   
  \caption{Mean range in $K$ magnitudes for models in the entire grid. 
  The SDDO values are plotted versus the SPL ones. Colours and symbols are explained 
  in Fig.~\ref{figDMMa1asKmagndus}.}
  \label{figDMMa1asKampl}
\end{figure}

We refer to the difference between maximum and minimum brightness during a period as the
amplitude, while the difference during all the computed epochs is referred to as the range.
We find that the range in $K$ can be larger by up to 0.5 
magnitudes for the highest mass-loss SDDO models with much dust in their winds (see Fig.~\ref{figDMMa1asKampl}).
Also, two episodic C5 models show large ranges in the SDDO case due to 
strong epoch-to-epoch variations.

\begin{figure}
  \resizebox{\hsize}{!}{\includegraphics{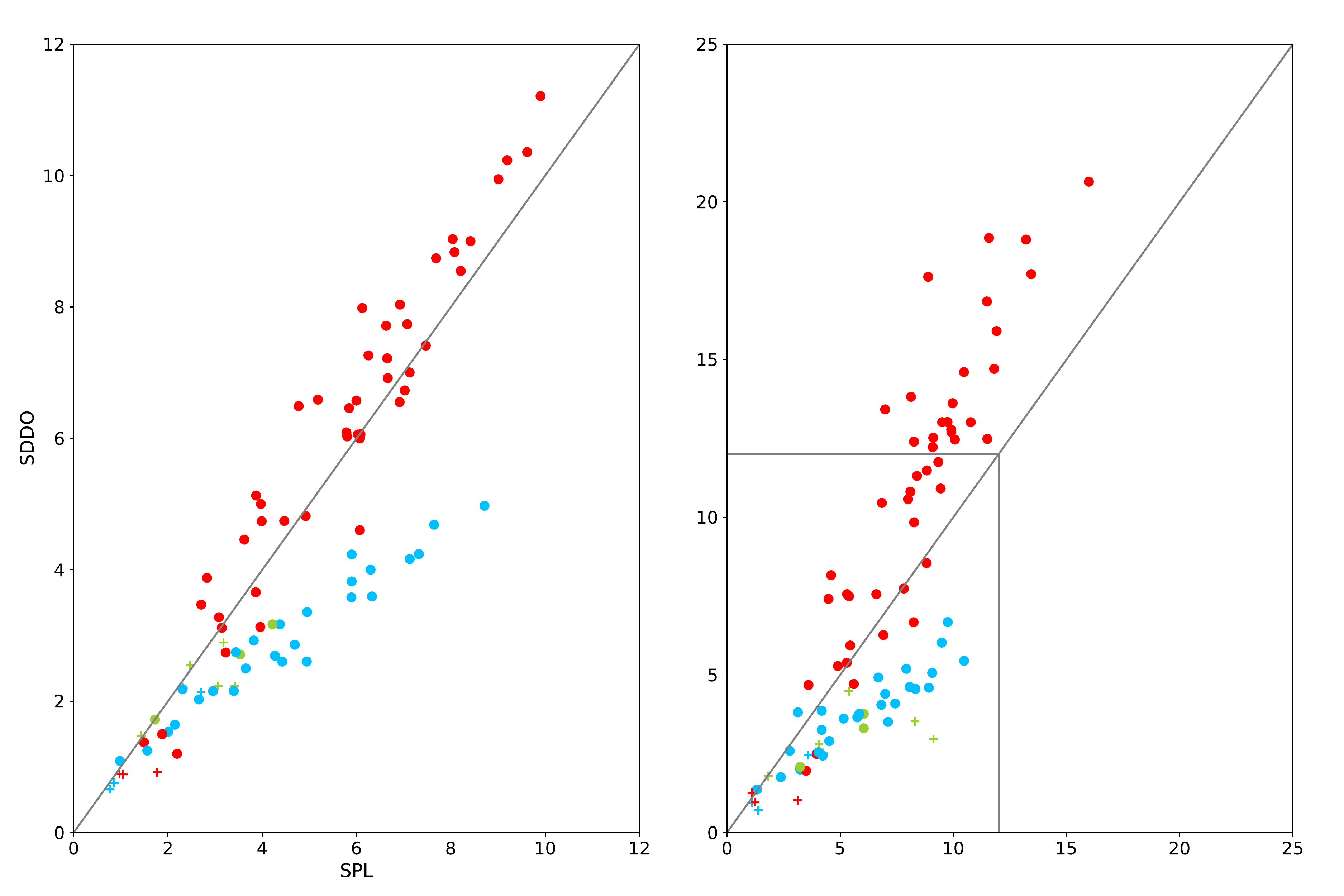}}
  \caption{Mean amplitude in $V$ magnitudes (left) and their range (right) 
  for models in the entire grid. The SDDO values are plotted versus the SPL ones. 
  The models are colour-coded according to their carbon excess (see Fig.~\ref{figDMMa1asKmagndus}).
  In the right panel, the area of the left panel is outlined.
  }
  \label{figDMMa1asVrnge}
\end{figure}

Plotting the $V$ magnitude amplitudes and ranges for the SDDO versus the SPL grid 
(Fig.~\ref{figDMMa1asVrnge}),
we see the same pattern as for the mean $V$ magnitude itself, as discussed above: 
smaller amplitude or range and a brighter mean value for moderate mass-losing models, 
while models with heavy mass loss show larger variations and a fainter mean value when the
SPL assumption is replaced by SDDO. 
Comparing the left and right panels in Fig.~\ref{figDMMa1asVrnge},
some models show much larger ranges in long-term variation than the amplitude over
a period; this reflects gusts of dust
occurring now and then, extending the range of variations in $V$.
We also note, as we show in Fig.~\ref{figDMMa1as_mag}, that the $V$ magnitudes are smaller (brighter)
for the lower carbon excesses and the C6 models and have smaller amplitudes.

A further examination of the results of the radiative transfer calculations done 
with {\tt COMA} is provided in the appendix.
The optical depth  as a function of distance to the 
stellar surface, for a maximum phase in the model in Fig.~\ref{figDMM_lc2c_76},
are  shown in Fig.~\ref{App_tau_C7u6_sddo_10} and
Fig.~\ref{App_tau_C7u6_spl_10} for SDDO and SPL, respectively.
We draw the reader's attention to how far from the star the light in for instance 
the $V$ band originates for models with a large carbon excess.


\section{Comparisons to observational data}

\begin{figure}
  \resizebox{1.1\hsize}{!}{\includegraphics{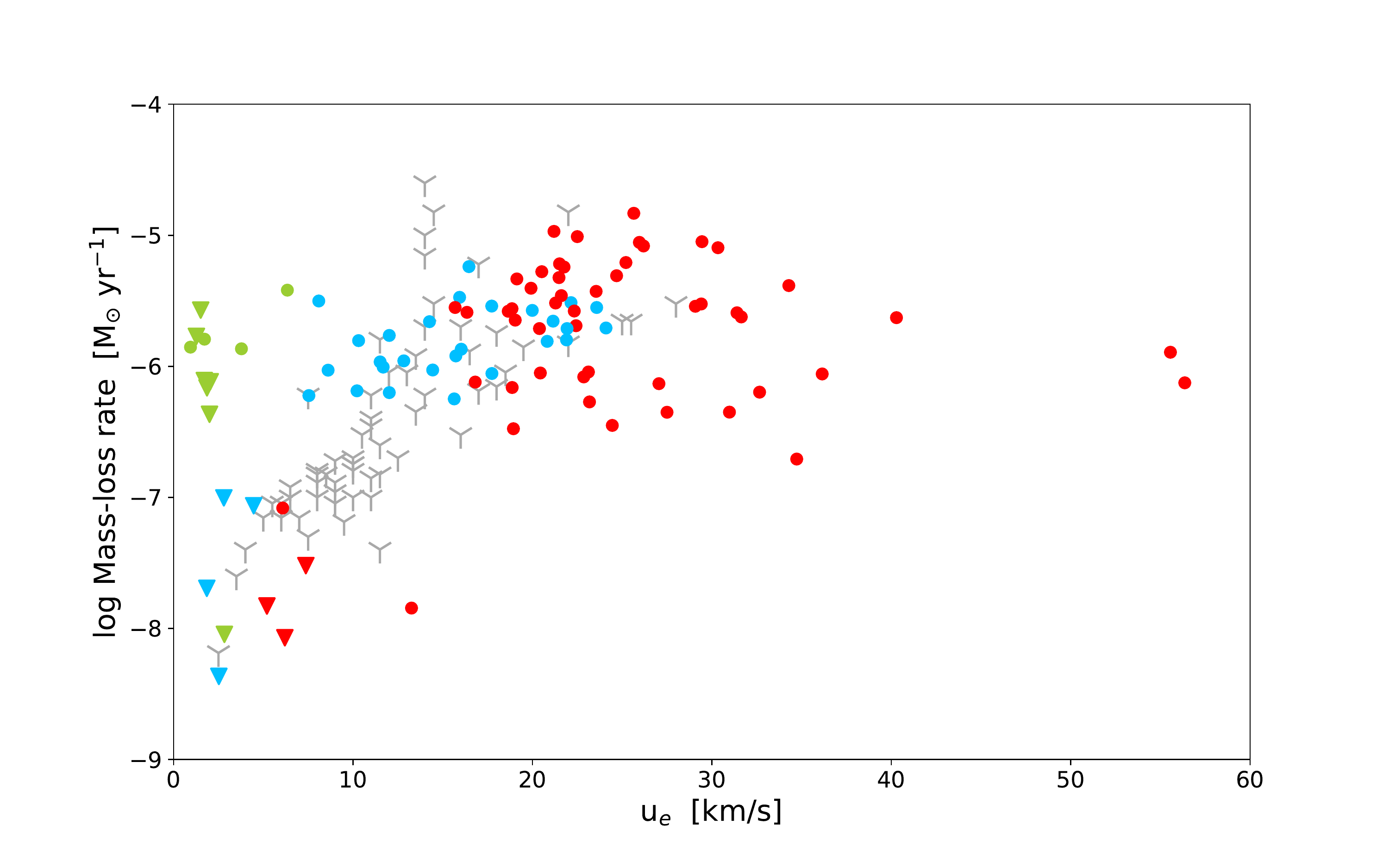}}   
  \caption{Mass-loss rates versus outflow speeds for wind models in the grid for the SDDO case. 
  The different colours denote the carbon excess. 
  Colours and symbols are explained in Fig.~\ref{figDMMa1asKmagndus}.  
  Observational data from \citet{SO01}  and \cite{RO14} are plotted as grey symbols.}
  \label{DMMmdotuext}
\end{figure}

In this section, we compare the new results to several observational studies of carbon stars. 
One useful way of displaying the dynamic properties of a grid is to plot  
the mass-loss rates versus the outflow speeds. 
Both quantities can be estimated from millimetre-wave observations of CO rotational lines and 
radiative transfer modelling of these lines.
For the SDDO models, such a diagram is shown in Fig.~\ref{DMMmdotuext}, and this can be 
compared to Fig.~4 in \cite{E+14} for the SPL case.
As noted in Sect.~\ref{s:DMAdyndust}, these quantities do not change much 
when SPL is replaced by SDDO.
In Fig.~\ref{DMMmdotuext}, we display these quantities from the present grid and observations 
of carbon stars given by
\citet{SO01} and \citet{RO14} with updated values for sources in common.
We note a general agreement, but find some models with very high outflow speeds 
not seen in these observations.
Also, some models with low outflow speeds and rather high mass-loss rates 
lack observed counterparts.
We highlight, however, that not all parameter combinations in 
the computed grid necessarily occur in nature, or they may be very rare 
or just not yet observed.
In summary, it appears that the models with a moderate carbon excess (C6) show a trend
and values of mass-loss rates and wind velocities that agree best with the observations.

\begin{figure}    
  \resizebox{1.1\hsize}{!}{\includegraphics{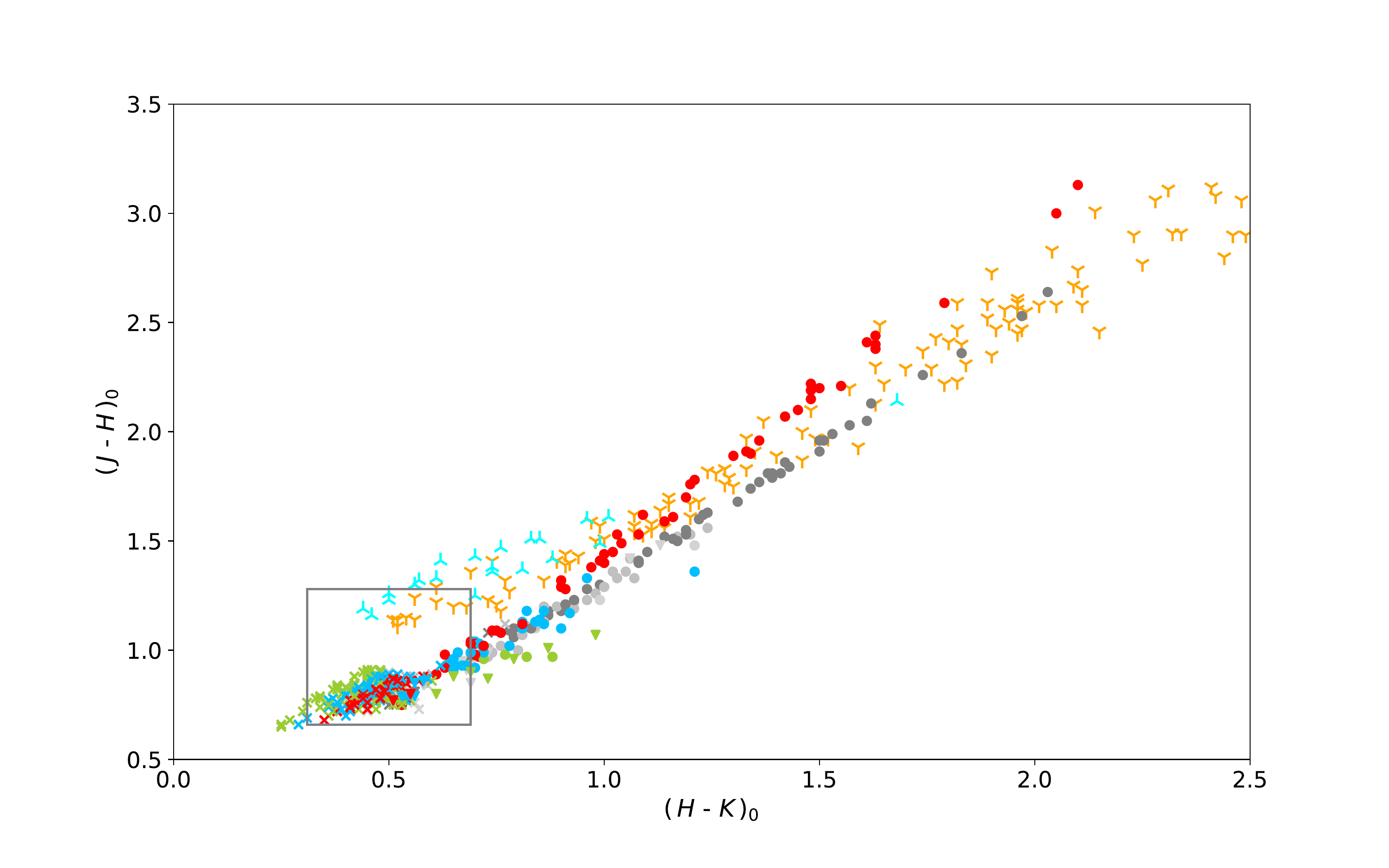}}   
  \caption{Mean \mbox{($J$\,--\,$H$)} colour versus mean \mbox{($H$\,--\,$K$)} colour
  for models in the entire grid. 
  Models using SPL are plotted in grey tones, and those with SDDO are coloured according 
  to carbon excess (see Fig.~\ref{figDMMa1asKmagndus}). 
  Observed carbon stars from \citet{W+06} are also plotted: orange symbols for 
  Miras and cyan for non-Mira stars. The grey box centered at (0.5, 1.0) outlines the area 
  in Fig.~\ref{figpipebowl}.}
  \label{figDMMa1asJHHK}
\end{figure}

\begin{figure}    
  \resizebox{1.0\hsize}{!}{\includegraphics{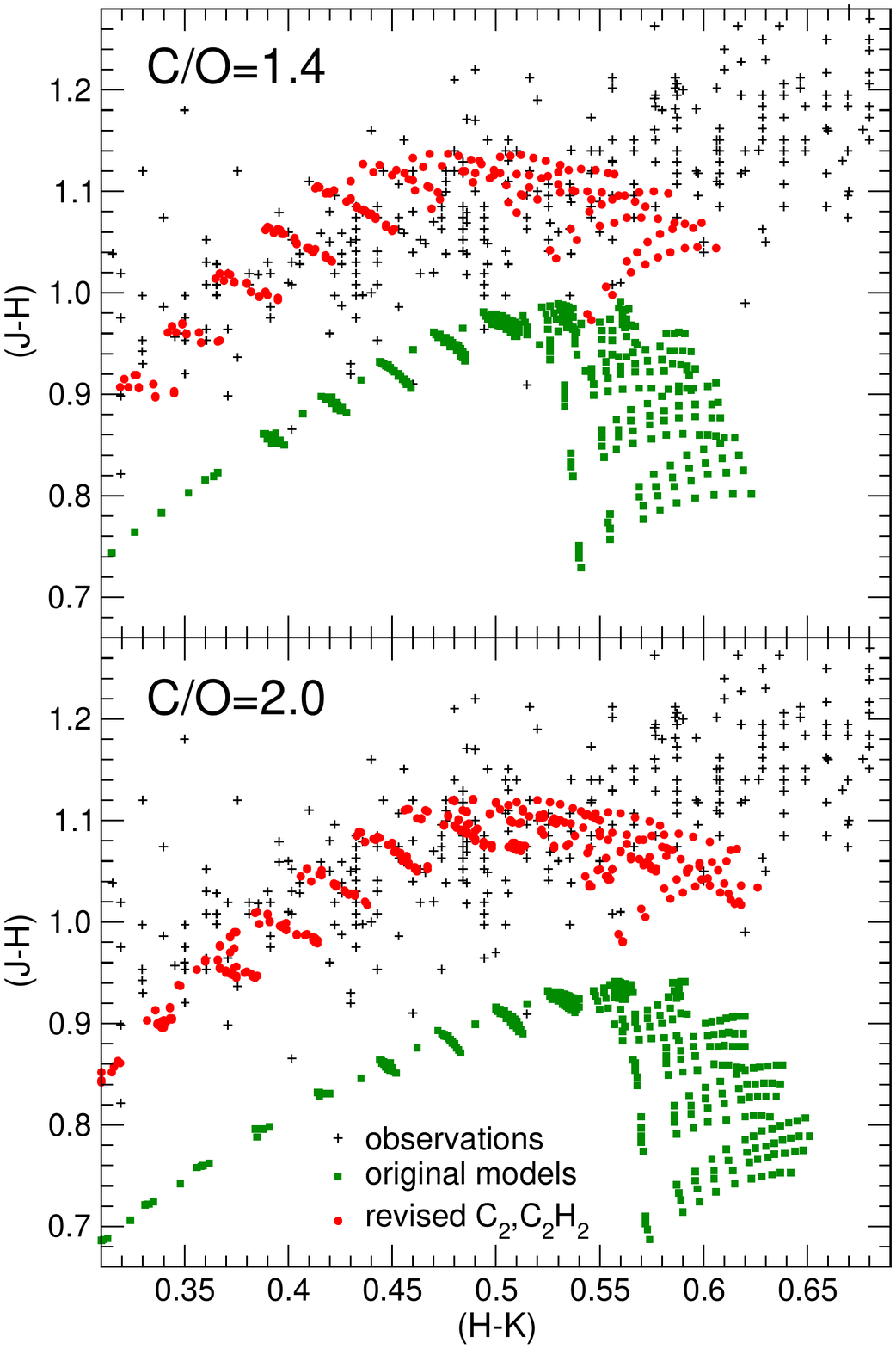}}   
  \caption{\mbox{($J$\,--\,$H$)} versus \mbox{($H$\,--\,$K$)} for hydrostatic dust-free 
  carbon star models from the COMARCS grid presented in 
  \citet[][green squares]{Aring+16, Aring+19}.
  Two C/O ratios are shown: 1.4 and 2.0. 
  The results are compared to a similar grid computed with updated opacities for molecular 
  and atomic transitions (red circles). 
  The large shift between the two sets of models is due to changes of the absorption caused 
  by C$_2$ and C$_2$H$_2$. In addition, observed values of carbon stars are included 
  (crosses); they 
  cover objects with solar and subsolar metallicities (see text).
  }
  \label{figpipebowl}
\end{figure}

The two-colour diagram, \mbox{($J$\,--\,$H$)} versus \mbox{($H$\,--\,$K$),} is commonly used
to compare models to observations. 
In Fig.~\ref{figDMMa1asJHHK}, we show this diagram 
for the SDDO models using coloured symbols and for the SPL ones 
\citep[originally shown in Fig.~13 in][]{E+14} in grey tones,
together with observational data from \citet{W+06}.
The slope in the diagram is steeper for the SDDO grid than for the SPL one.
This is mostly due to \mbox{($J$\,--\,$H$)} becoming larger for SDDO models: 
$H$ gets brighter more than $J$ does (if at all; 
see Sect.~\ref{s:DMAphot}, as well as Appendix~\ref{App_A}).
Thus, we note that this slope depends on the details of the dust opacities.

Comparing the SDDO grid to 
the photometric near-infrared observations by \citet{W+06},
we used the data in their Table~6 where de-reddened $J$, $H$, and $K$ magnitudes are listed. 
Also, some data from their Table~3 are included in our comparison, 
namely for sources with $K < 4,$ which can be assumed not to be significantly reddened
by interstellar extinction.
The colours for the SDDO grid agree with observations rather well for 
1.0 $\lesssim$ \mbox{($H$\,--\,$K$)} $\lesssim$ 1.5 (i.e. the range of moderately dusty winds).
For \mbox{($H$\,--\,$K$)} $\gtrsim$ 1.5, we find SDDO models with low effective temperatures,
high luminosities, large carbon excess (C7) and high mass-loss rates, typically 
10$^{-5}$ M$_\odot$\,yr$^{-1}$, very red \mbox{($J$\,--\,$K$)} colours ($\gtrsim$ 4), and 
high levels of dust (gas-to-dust ratios $\sim$ 500--800; see below). 
Different variability classes for the observed stars are plotted with different colours.
The non-Mira stars in \citet{W+06} fit less well, especially for 
\mbox{($H$\,--\,$K$)} $\lesssim$ 1.0; in this interval, updated gas opacities for the
models could improve the agreement.

As an example, in Fig.~\ref{figpipebowl} we show a \mbox{($J$\,--\,$H$)} versus \mbox{($H$\,--\,$K$)}
diagram for dust-free {\sl \emph{hydrostatic}} carbon star models from \citet[][]{Aring+16, Aring+19}.
The models are computed with the COMARCS code and a have solar metallicity of
$\rm log(g~[cm/s^2]) \le 0.0$, effective temperatures down to 2500~K, and different masses.
Two different C/O values are used: 1.4 and 2.0 (here, only the abundance of carbon is changed with
respect to a solar mixture). For each C/O value, two sets of models are shown: the original
ones and a set with updated gas opacities. For C$_2,$ data by \citet{Quer+74} is replaced by
\citet[][]{Yurch+18, McKe+20}, and for C$_2$H$_2,$ data from \citet{Jorg+89} is replaced by
\citet{Chubb+20}. 
The large differences between the two sets of models are due to differences of these gas opacities. 
In the figure, we also plot observational values for carbon stars selected to be 
little influenced by dust.
The observational values come from \citet{Berg+01, Cohen+81, Costa+96, Gonn+16,
Tott+00, GGCGL12} and agree much better with the set of hydrostatic dust-free models
with the updated gas opacities.

The grey box in Fig.~\ref{figDMMa1asJHHK} centred at (0.5, 1.0) outlines the ranges 
covered by Fig.~\ref{figpipebowl},
indicating that the inclusion of updated C$_2$ and C$_2$H$_2$ gas opacities
in future hydrodynamic DARWIN models should improve the agreement with
observations for nearly dust-free models.
In the current paper, focussing on dust opacities, however, we chose to
keep the same set of gas opacity data as in \citet{E+14} to make the SPL and SDDO
grids comparable in that respect.

\begin{figure}
  \resizebox{1.1\hsize}{!}{\includegraphics{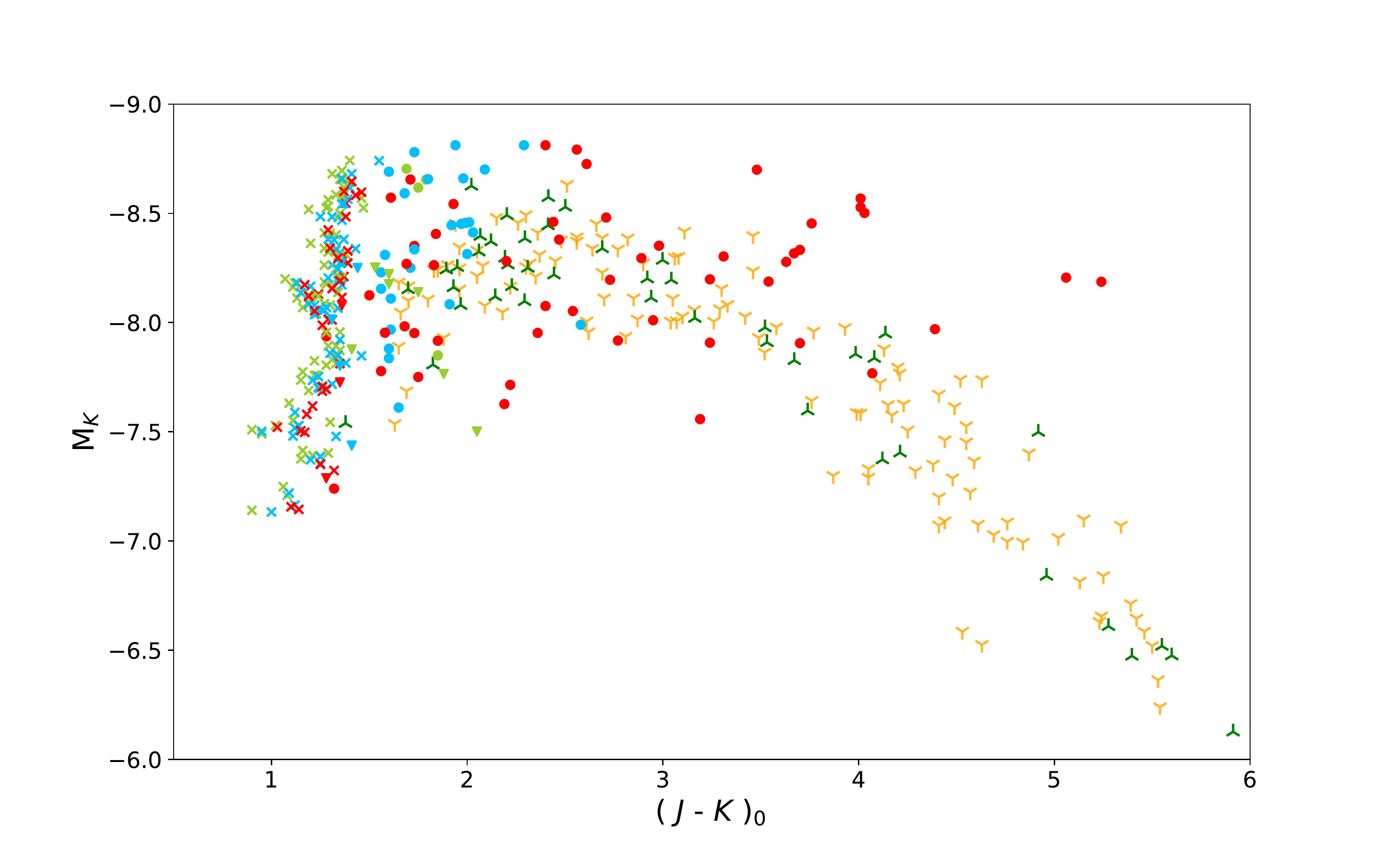}}   
  \caption{M$_K$ values versus mean \mbox{($J$\,--\,$K$)}$_0$ colour for SDDO 
  models in the entire grid. 
  Colours and symbols are explained in Fig.~\ref{figDMMa1asKmagndus}.
  Observed carbon Mira stars from \citet{W+06} are plotted as orange symbols, 
  and those from \citet{Menz+06} are plotted as green symbols.}
  \label{figDMM_MK_JmK}
\end{figure}

In the colour-magnitude diagram (Fig.~\ref{figDMM_MK_JmK}) displaying M$_K$ versus 
\mbox{($J$\,--\,$K$)$_0,$} the observed values from \citet{W+06}, Table~6, come from de-reddened 
magnitudes and a bolometric PL relation for LMC to derive distances and bolometric corrections. 
For northern sources, we used the values in \citet{Menz+06}, Table~3, 
to calculate M$_K$ and \mbox{($J$\,--\,$K$)$_0$}. 
We see that the values from our SDDO grid bracket the observed values for 
\mbox{($J$\,--\,$K$)$_0$} $\lesssim$ 4, while the redder stars seem to be affected by larger 
circumstellar extinction in $K$ than the models in our grid.

\begin{figure}
  \resizebox{1.1\hsize}{!}{\includegraphics{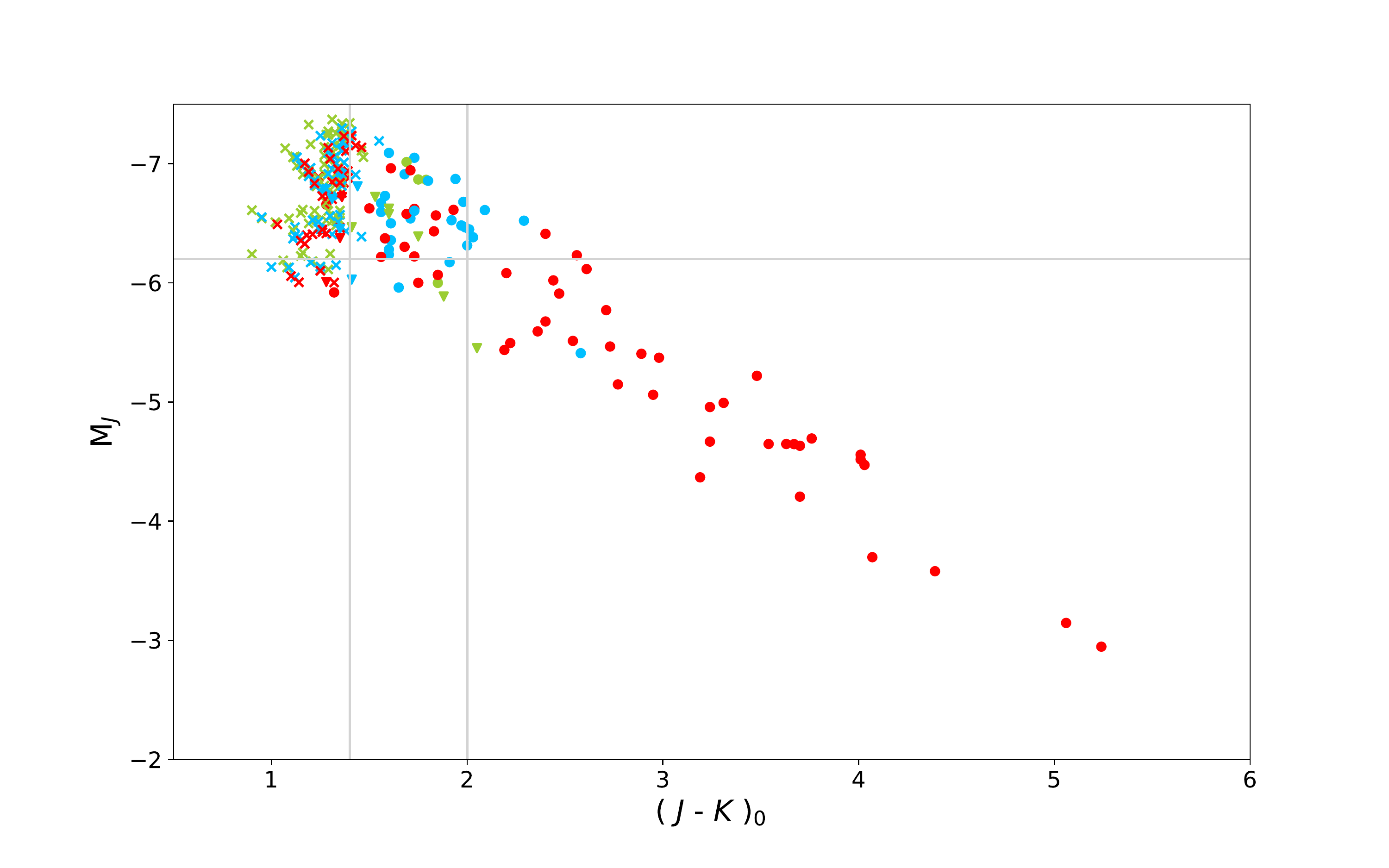}}   
  \caption{M$_J$ values versus mean \mbox{($J$\,--\,$K$)}$_0$ colour for the SDDO 
  models in the entire grid. 
  Colours and symbols are explained in Fig.~\ref{figDMMa1asKmagndus}.
  The value of M$_J$ = -6.2 is marked, as well as \mbox{($J$\,--\,$K$)}$_0$ = 1.4 and 2.0.
  }
  \label{fig_MJ_JmK}
\end{figure}

The distribution of carbon stars in colour-magnitude dia\-grams is relatively flat, which makes them
potential candidates as standard candles for (extragalactic) distance determinations, especially
as they are intrinsically very bright.
This was first studied by \citet{Richer+84}, who used VRI photometry to estimate the distance to NGC\,205, 
and later by \citet{WN01}, who calibrated the M$_J$ versus \mbox{($J$\,--\,$K_s$)} distribution of carbon stars
in a narrow range in \mbox{($J$\,--\,$K_s$)} from 2MASS photometry of the Magellanic Clouds.
Recently, this so-called JAGB method has been used for distance determinations by \citet{Ripoche+20}
and simultaneously by \citet{MadoreFreedman20} and in later publications by these groups.
Here stars in a population with 1.4 < \mbox{($J$\,--\,$K_s$)} < 2.0 are assumed to be carbon stars
with a well-determined M$_J$ of -6.2.
The precision of this method depends on selection effects in the samples of calibrators, for example the
metallicity \citep[as discussed by][]{Lee+21}. In Fig.~\ref{fig_MJ_JmK}, we show the mean M$_J$ versus
\mbox{($J$\,--\,$K_s$)$_0$} for the SDDO grid, and we note a general similarity to Fig.~1 of
\citet{Madore+22}. In all magnitude-limited samples, one suffers from a Malmquist bias, and
\citet{Parada+21} discussed the advantage of using median instead of mean magnitudes for stars
in extragalactic surveys. There might also be a difference in our grid between the median and mean
magnitudes derived from the light curves. 
As presented in Appendix \ref{a:medianmean}, we find that the median magnitude is generally
somewhat brighter than the mean one, but this effect is only significant for some models with
the highest carbon excesses (C7).
This is due to asymmetric light curves (an effect of very gusty winds).
The effect on the $J$ magnitude is mostly much smaller than 0.2 magnitudes.
For reasons of comparability with earlier papers and with the SPL grid, we therefore used mean
magnitudes for the SDDO models in the figures discussed in this section.

\begin{figure}
  \resizebox{1.1\hsize}{!}{\includegraphics{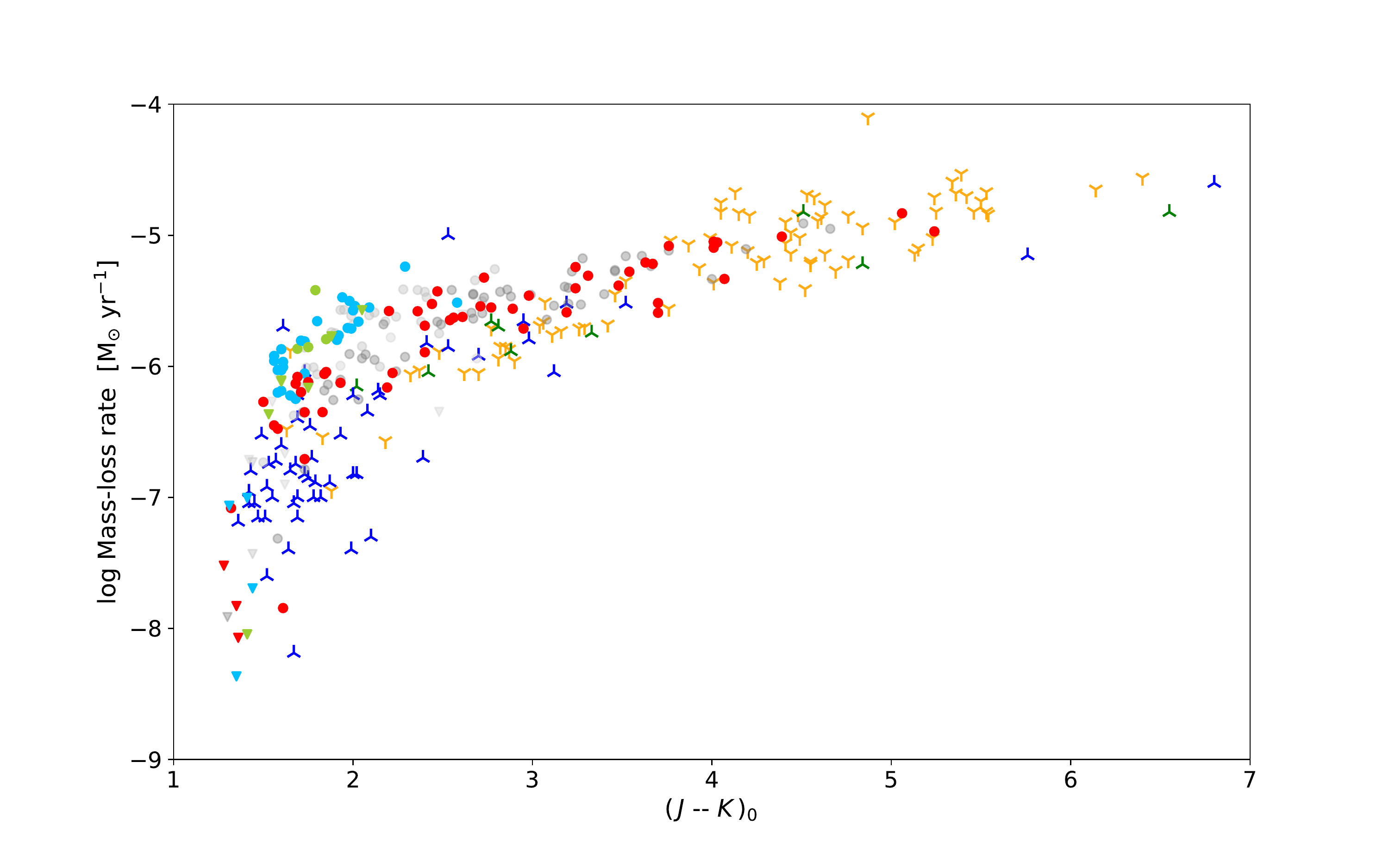}}  
  \caption{Mass-loss rate versus mean \mbox{($J$\,--\,$K$)$_0$} colour
  for the SDDO wind models in the grid. 
  The colours denote the carbon excess (see Fig.~\ref{figDMMa1asKmagndus}), the circles 
  with different grey hues are the SPL models. 
  Observed carbon Mira stars from \citet{W+06} are plotted as
  orange symbols, while observations from \citet{SO01} and \cite{RO14} are plotted as 
  blue symbols \citep[or green if the source is also present as 
  an orange symbol from][]{W+06}.}
  \label{figDMM_mdot_JmK}
\end{figure}
\begin{figure}
  \resizebox{1.1\hsize}{!}{\includegraphics{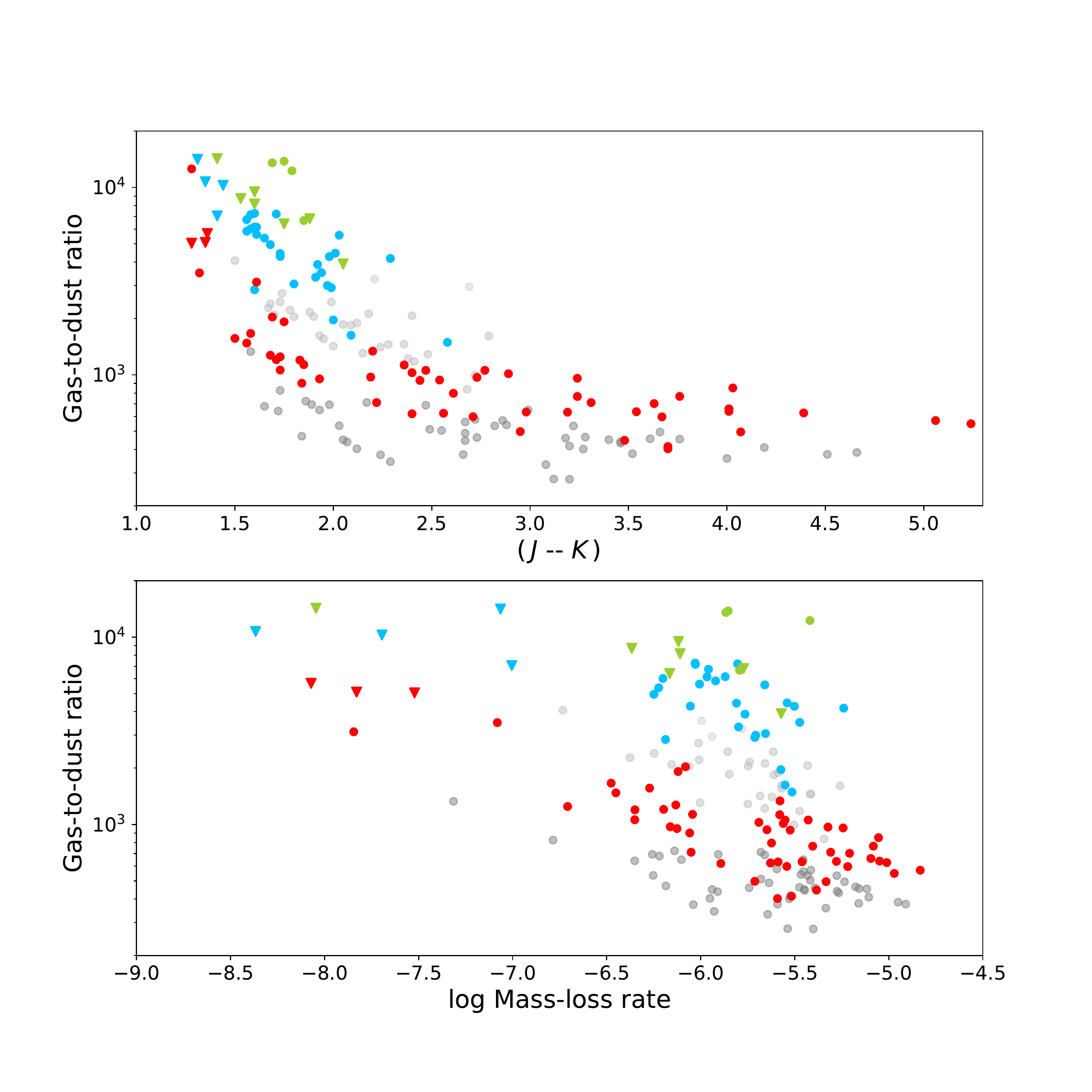}}   
  \caption{Mean gas-to-dust ratio for the wind models in the grid plotted against
  the mean \mbox{($J$\,--\,$K$)} colour (top panel) and against 
  the mean mass-loss rate (lower panel). 
  The models are colour-coded for the carbon excess (see Fig.~\ref{figDMMa1asKmagndus}). For
  models computed with the SPL, we use grey symbols.  }
  \label{FigDMM_g2d}
\end{figure}

\begin{figure}
   \resizebox{1.1\hsize}{!}{\includegraphics{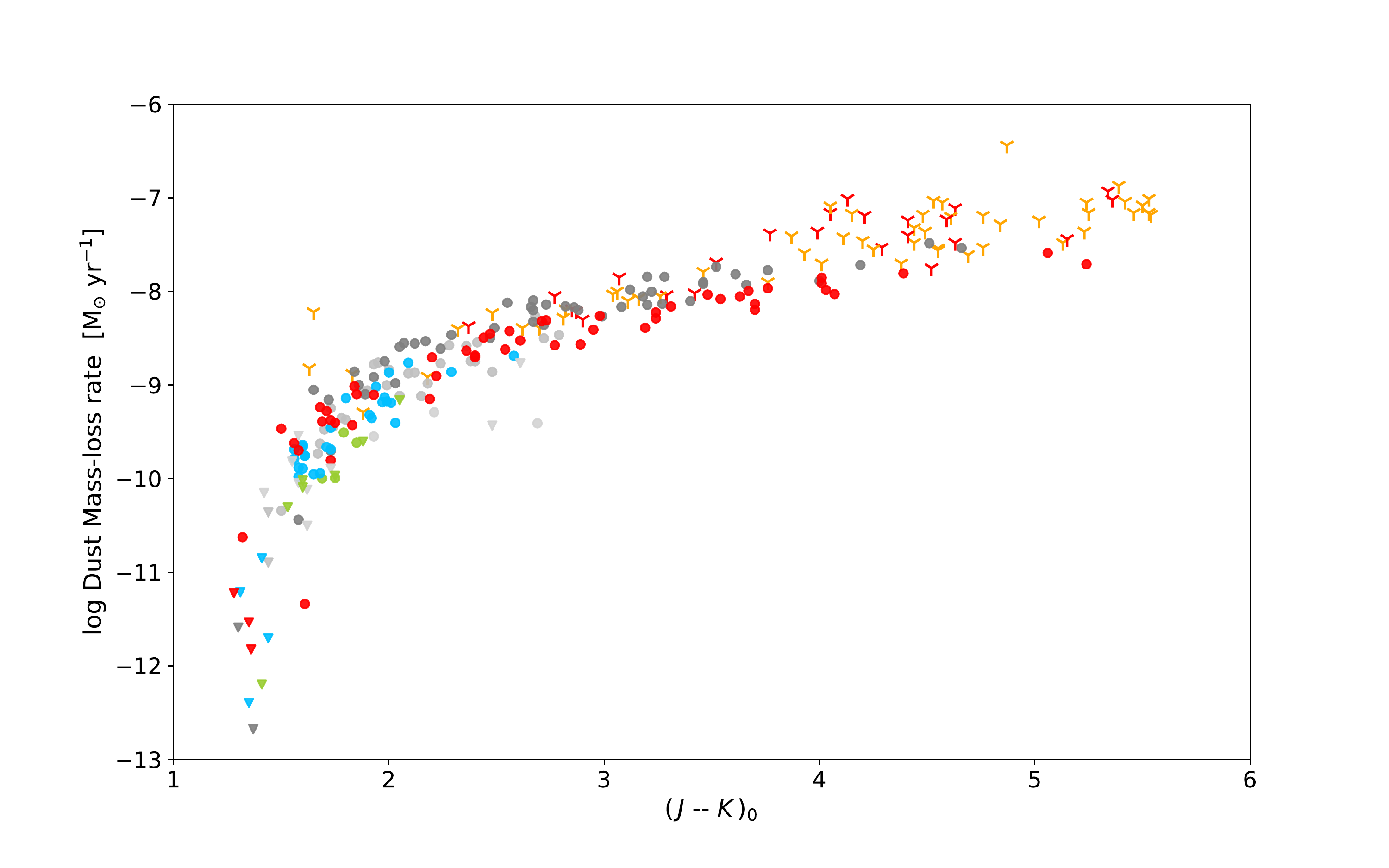}}
   \caption{Mean dust-mass-loss rate versus mean \mbox{($J$\,--\,$K$)$_0$} 
   colour for the wind models in the grid. 
   The observed dust-mass-loss rates (plotted as orange symbols) were derived
   from total mass-loss rates in \citet{W+06} using a gas-to-dust ratio of 220.}
   \label{figDMM_dust_mdot_JmK}
\end{figure}

The mass-loss rates for the SDDO models are plotted as a function of the 
\mbox{($J$\,--\,$K$)$_0$} colour in Fig.~\ref{figDMM_mdot_JmK}.
For comparison,  corresponding values given by \citet{W+06} are shown. 
These are
derived from the 60 $\mu$m IRAS fluxes (assumed to originate from dust in the
circumstellar envelopes) using an expression from \citet{Jura87}.
Mass-loss rates by \citet{SO01} and \citet{RO14} from CO data are also plotted.
We find a reasonable general agreement between models and observations
given the underlying uncertainties. 
In particular,
when comparing with the mass-loss rates given by \citet{W+06}, we must 
remember that these values are based on \citet{Jura87}, who used a gas-to-dust 
ratio of  220 to convert the dust-mass-loss rates into total mass loss.
This ratio is considerably lower than in our models (see Fig.~\ref{FigDMM_g2d}).  
We find values of the order of 10$^3$  for our grid of models,
reflecting typical condensation degrees well below 1 (see Fig.~\ref{figDMMa1s}).
These gas-to-dust values are considerably higher than those often assumed in the literature.

Taking the mass-loss rates in \citet{W+06} and their value for
the gas-to-dust ratio adopted from \citet{Jura87}, we calculated the dust-mass-loss rates in their
sample. 
These are plotted in Fig.~\ref{figDMM_dust_mdot_JmK}, together with the dust-mass-loss rates
for the SDDO grid.
We note a better agreement between model values and those derived from observations than 
for the total mass-loss rates (Fig.~\ref{figDMM_mdot_JmK}). 
For \mbox{($J$\,--\,$K$)} $\gtrsim$ 4, however, the models tend to have lower 
mass-loss rates at a given \mbox{($J$\,--\,$K$)}.
This is interesting in light of the discrepancies in the colour-magnitude diagram
(Fig.~\ref{figDMM_MK_JmK}) in the same \mbox{($J$\,--\,$K$)} range.


\section{Summary and conclusions}

Carbonaceous dust is crucial for driving the stellar winds in carbon stars, and the mass loss
will eventually terminate their evolution along the AGB. 
We computed grids of dynamical atmosphere and wind models representing these stars in order
to investigate the effects of changing the dust opacity from the simple 
small particle approximation (SPL) to a size-dependent treatment using Mie calculations. 
The grids include 268 combinations of physical parameters appropriate 
for C-rich AGB stars (effective temperature, luminosity, stellar mass, carbon excess, and
pulsational properties). 
Mass loss (stellar winds) developed in about 1/3 of our 
radiation-hydrodynamic DARWIN models, which include time-dependent dust 
formation/sublimation and frequency-dependent radiative transfer. 
One of the main conclusions of comparing the two grids is that the mass-loss rates and outflow 
speeds are not strongly affected by replacing the SPL approximation with SDDOs.
Due to a higher radiative force on sub-micron-sized dust grains (see Fig.~\ref{f_Q_V_agr}), 
compared to the SPL case, they travel faster through the dust-forming outer atmosphere layers, 
which results in smaller dust grains and lower condensation degrees 
(and thus higher gas-to-dust ratios). 
The dynamics is tending towards more narrow and denser shells of mass loss, a more 
strongly variable behaviour of the winds in the new SDDO models. 

We also calculated detailed {\it \emph{a posteriori}} radiative transfer through 
the spherically symmetric radial structures for hundreds of snapshots per model 
using the COMA code. 
The resulting light curves and colour-colour diagrams, derived from the COMA flux distributions, 
show several differences between the SPL and the size-dependent opacity cases. 
Compared to SPL, we find generally brighter K magnitudes; for shorter wavelengths where the dust
influence is 
more complex, we find brighter magnitudes when the dust absorption is 
small and dimmer magnitudes with large absorption, especially in the $V$ band
(all compared to the SPL case). 

Some comparisons to observational data have been made, for example mass-loss rates versus outflow speeds.
In Fig.~\ref{DMMmdotuext},
we compare model values to results from CO rotational line observations.
Also, synthetic photometric mean magnitudes are compared to NIR observations in 
Fig.~\ref{figDMMa1asJHHK}.
Both these comparisons show a general agreement, but we must remember that all parameter 
combinations used in the model have equal weight and do not represent 
an observed carbon star population. 
Other laboratory data for the optical properties of amorphous carbon as well as the effects of
using updated gas opacities should be explored in the future.
We will also extend our investigations to additional
photometric systems. This will primarily cover the Gaia-2MASS
diagram used to identify carbon stars \citep[e.g.][]{Abia+22,Lebzelter+18}
and mid-infrared
data from different space missions (Spitzer, WISE, Akari), which contain information
about the cool dust in the outer envelopes \citep[e.g.][]{Srinivasan+16, Nanni+18, Nanni+19}.

\begin{acknowledgements}
SH acknowledges funding from the European Research Council (ERC) under the European Union's 
Horizon 2020 research and innovation programme (Grant agreement No. 883867, project EXWINGS)
and the Swedish Research Council (Vetenskapsr\aa det, grant number 2019-04059). 
The computations were enabled by resources provided by the Swedish National 
Infrastructure for Computing (SNIC) at UPPMAX partially funded by the 
Swedish Research Council through grant agreement no. 2018-05973. 

\end{acknowledgements}

%
\bibliographystyle{aa} 
\bibliography{DMA20} 

\begin{thebibliography}{58}
\expandafter\ifx\csname natexlab\endcsname\relax\def\natexlab#1{#1}\fi

\bibitem[{{Abia} {et~al.}(2022){Abia}, {de Laverny}, {Romero-G{\'o}mez}, \&
  {Figueras}}]{Abia+22}
{Abia}, C., {de Laverny}, P., {Romero-G{\'o}mez}, M., \& {Figueras}, F. 2022,
  \aap, 664, A45

\bibitem[{Aringer(2000)}]{Aring00}
Aringer, B. 2000, PhD thesis, Univ. of Vienna

\bibitem[{{Aringer} {et~al.}(2016){Aringer}, {Girardi}, {Nowotny}, {Marigo}, \&
  {Bressan}}]{Aring+16}
{Aringer}, B., {Girardi}, L., {Nowotny}, W., {Marigo}, P., \& {Bressan}, A.
  2016, \mnras, 457, 3611

\bibitem[{{Aringer} {et~al.}(2009){Aringer}, {Girardi}, {Nowotny}, {Marigo}, \&
  {Lederer}}]{AGNML09}
{Aringer}, B., {Girardi}, L., {Nowotny}, W., {Marigo}, P., \& {Lederer}, M.~T.
  2009, \aap, 503, 913

\bibitem[{{Aringer} {et~al.}(2019){Aringer}, {Marigo}, {Nowotny}, {Girardi},
  {Me{\v{c}}ina}, \& {Nanni}}]{Aring+19}
{Aringer}, B., {Marigo}, P., {Nowotny}, W., {et~al.} 2019, \mnras, 487, 2133

\bibitem[{{Asplund} {et~al.}(2005){Asplund}, {Grevesse}, \& {Sauval}}]{AsGS05}
{Asplund}, M., {Grevesse}, N., \& {Sauval}, A.~J. 2005, in Astronomical Society
  of the Pacific Conference Series, Vol. 336, Cosmic Abundances as Records of
  Stellar Evolution and Nucleosynthesis, ed. T.~G. {Barnes}, III \& F.~N.
  {Bash}, 25

\bibitem[{{Bergeat} {et~al.}(2001){Bergeat}, {Knapik}, \& {Rutily}}]{Berg+01}
{Bergeat}, J., {Knapik}, A., \& {Rutily}, B. 2001, \aap, 369, 178

\bibitem[{{Bessell}(1990)}]{Besse90}
{Bessell}, M.~S. 1990, \pasp, 102, 1181

\bibitem[{{Bessell} \& {Brett}(1988)}]{BB88}
{Bessell}, M.~S. \& {Brett}, J.~M. 1988, \pasp, 100, 1134

\bibitem[{{Bladh} {et~al.}(2019){Bladh}, {Eriksson}, {Marigo}, {Liljegren}, \&
  {Aringer}}]{Bladh+19}
{Bladh}, S., {Eriksson}, K., {Marigo}, P., {Liljegren}, S., \& {Aringer}, B.
  2019, \aap, 623, A119

\bibitem[{{Chubb} {et~al.}(2020){Chubb}, {Tennyson}, \& {Yurchenko}}]{Chubb+20}
{Chubb}, K.~L., {Tennyson}, J., \& {Yurchenko}, S.~N. 2020, \mnras, 493, 1531

\bibitem[{{Cohen} {et~al.}(1981){Cohen}, {Frogel}, {Persson}, \&
  {Elias}}]{Cohen+81}
{Cohen}, J.~G., {Frogel}, J.~A., {Persson}, S.~E., \& {Elias}, J.~H. 1981,
  \apj, 249, 481

\bibitem[{{Costa} \& {Frogel}(1996)}]{Costa+96}
{Costa}, E. \& {Frogel}, J.~A. 1996, \aj, 112, 2607

\bibitem[{{Eriksson} {et~al.}(2014){Eriksson}, {Nowotny}, {H{\"o}fner},
  {Aringer}, \& {Wachter}}]{E+14}
{Eriksson}, K., {Nowotny}, W., {H{\"o}fner}, S., {Aringer}, B., \& {Wachter},
  A. 2014, \aap, 566, A95

\bibitem[{{Feast} {et~al.}(1989){Feast}, {Glass}, {Whitelock}, \&
  {Catchpole}}]{FeGWC89}
{Feast}, M.~W., {Glass}, I.~S., {Whitelock}, P.~A., \& {Catchpole}, R.~M. 1989,
  \mnras, 241, 375

\bibitem[{{Freedman} \& {Madore}(2020)}]{MadoreFreedman20}
{Freedman}, W.~L. \& {Madore}, B.~F. 2020, \apj, 899, 67

\bibitem[{{Gonneau} {et~al.}(2016){Gonneau}, {Lan{\c{c}}on}, {Trager},
  {Aringer}, {Lyubenova}, {Nowotny}, {Peletier}, {Prugniel}, {Chen}, {Dries},
  {Choudhury}, {Falc{\'o}n-Barroso}, {Koleva}, {Meneses-Goytia},
  {S{\'a}nchez-Bl{\'a}zquez}, \& {Vazdekis}}]{Gonn+16}
{Gonneau}, A., {Lan{\c{c}}on}, A., {Trager}, S.~C., {et~al.} 2016, \aap, 589,
  A36

\bibitem[{{Groenewegen} {et~al.}(2020){Groenewegen}, {Nanni}, {Cioni},
  {Girardi}, {de Grijs}, {Ivanov}, {Marconi}, {Moretti}, {Oliveira},
  {Petr-Gotzens}, {Ripepi}, \& {van Loon}}]{Groenw+20}
{Groenewegen}, M.~A.~T., {Nanni}, A., {Cioni}, M. R.~L., {et~al.} 2020, \aap,
  636, A48

\bibitem[{{Gullieuszik} {et~al.}(2012){Gullieuszik}, {Groenewegen}, {Cioni},
  {de Grijs}, {van Loon}, {Girardi}, {Ivanov}, {Oliveira}, {Emerson}, \&
  {Guandalini}}]{GGCGL12}
{Gullieuszik}, M., {Groenewegen}, M.~A.~T., {Cioni}, M.-R.~L., {et~al.} 2012,
  \aap, 537, A105

\bibitem[{{Herwig}(2005)}]{Herw05}
{Herwig}, F. 2005, \araa, 43, 435

\bibitem[{{H{\"o}fner} {et~al.}(2016){H{\"o}fner}, {Bladh}, {Aringer}, \&
  {Ahuja}}]{hoefetal16}
{H{\"o}fner}, S., {Bladh}, S., {Aringer}, B., \& {Ahuja}, R. 2016, \aap, 594,
  A108

\bibitem[{{H{\"o}fner} {et~al.}(2003){H{\"o}fner}, {Gautschy-Loidl}, {Aringer},
  \& {J{\o}rgensen}}]{HoGAJ03}
{H{\"o}fner}, S., {Gautschy-Loidl}, R., {Aringer}, B., \& {J{\o}rgensen}, U.~G.
  2003, \aap, 399, 589

\bibitem[{{H{\"o}fner} \& {Olofsson}(2018)}]{HoOl18}
{H{\"o}fner}, S. \& {Olofsson}, H. 2018, \aapr, 26, 1

\bibitem[{{Ita} {et~al.}(2004){Ita}, {Tanab{\'e}}, {Matsunaga}, {Nakajima},
  {Nagashima}, {Nagayama}, {Kato}, {Kurita}, {Nagata}, {Sato}, {Tamura},
  {Nakaya}, \& {Nakada}}]{Ita04}
{Ita}, Y., {Tanab{\'e}}, T., {Matsunaga}, N., {et~al.} 2004, \mnras, 353, 705

\bibitem[{{Jorgensen} {et~al.}(1989){Jorgensen}, {Alml{\"o}f}, \&
  {Siegbahn}}]{Jorg+89}
{Jorgensen}, U.~G., {Alml{\"o}f}, J., \& {Siegbahn}, P. E.~M. 1989, \apj, 343,
  554

\bibitem[{{Jura}(1987)}]{Jura87}
{Jura}, M. 1987, \apj, 313, 743

\bibitem[{{Khouri} {et~al.}(2016){Khouri}, {Maercker}, {Waters}, {Vlemmings},
  {Kervella}, {de Koter}, {Ginski}, {De Beck}, {Decin}, {Min}, {Dominik},
  {O'Gorman}, {Schmid}, {Lombaert}, \& {Lagadec}}]{Khou+16}
{Khouri}, T., {Maercker}, M., {Waters}, L.~B.~F.~M., {et~al.} 2016, \aap, 591,
  A70

\bibitem[{{Lebzelter} {et~al.}(2018){Lebzelter}, {Mowlavi}, {Marigo},
  {Pastorelli}, {Trabucchi}, {Wood}, \& {Lecoeur-Ta{\"\i}bi}}]{Lebzelter+18}
{Lebzelter}, T., {Mowlavi}, N., {Marigo}, P., {et~al.} 2018, \aap, 616, L13

\bibitem[{{Lee} {et~al.}(2021){Lee}, {Freedman}, {Madore}, {Owens}, \& {Sung
  Jang}}]{Lee+21}
{Lee}, A.~J., {Freedman}, W.~L., {Madore}, B.~F., {Owens}, K.~A., \& {Sung
  Jang}, I. 2021, \apj, 923, 157

\bibitem[{{Li} {et~al.}(2018){Li}, {Luo}, {Du}, {Zuo}, {Wang}, {Zhao}, {Jiang},
  {Zhang}, {Liu}, {Qin}, {Wang}, {Du}, {Guo}, {Wang}, {Han}, {Xiang}, {Huang},
  {Chen}, {Chen}, {Kong}, {Hou}, {Song}, {Wang}, {Wu}, {Zhang}, {Zhang},
  {Wang}, {Cao}, {Hou}, \& {Zhao}}]{Li+18}
{Li}, Y.-B., {Luo}, A.~L., {Du}, C.-D., {et~al.} 2018, \apjs, 234, 31

\bibitem[{{Madore} {et~al.}(2022){Madore}, {Freedman}, {Lee}, \&
  {Owens}}]{Madore+22}
{Madore}, B.~F., {Freedman}, W.~L., {Lee}, A.~J., \& {Owens}, K. 2022, \apj,
  938, 125

\bibitem[{{Mattsson} \& {H{\"o}fner}(2011)}]{MatH11}
{Mattsson}, L. \& {H{\"o}fner}, S. 2011, \aap, 533, A42

\bibitem[{{Mattsson} {et~al.}(2010){Mattsson}, {Wahlin}, \&
  {H{\"o}fner}}]{MatWH10}
{Mattsson}, L., {Wahlin}, R., \& {H{\"o}fner}, S. 2010, \aap, 509, A14

\bibitem[{{McKemmish} {et~al.}(2020){McKemmish}, {Syme}, {Borsovszky},
  {Yurchenko}, {Tennyson}, {Furtenbacher}, \& {Cs{\'a}sz{\'a}r}}]{McKe+20}
{McKemmish}, L.~K., {Syme}, A.-M., {Borsovszky}, J., {et~al.} 2020, \mnras,
  497, 1081

\bibitem[{{Menzies} {et~al.}(2006){Menzies}, {Feast}, \& {Whitelock}}]{Menz+06}
{Menzies}, J.~W., {Feast}, M.~W., \& {Whitelock}, P.~A. 2006, \mnras, 369, 783

\bibitem[{{Nanni} {et~al.}(2019){Nanni}, {Groenewegen}, {Aringer}, {Rubele},
  {Bressan}, {van Loon}, {Goldman}, \& {Boyer}}]{Nanni+19}
{Nanni}, A., {Groenewegen}, M. A.~T., {Aringer}, B., {et~al.} 2019, \mnras,
  487, 502

\bibitem[{{Nanni} {et~al.}(2018){Nanni}, {Marigo}, {Girardi}, {Rubele},
  {Bressan}, {Groenewegen}, {Pastorelli}, \& {Aringer}}]{Nanni+18}
{Nanni}, A., {Marigo}, P., {Girardi}, L., {et~al.} 2018, \mnras, 473, 5492

\bibitem[{{Nowotny} {et~al.}(2013){Nowotny}, {Aringer}, {H{\"o}fner}, \&
  {Eriksson}}]{NoAHE13}
{Nowotny}, W., {Aringer}, B., {H{\"o}fner}, S., \& {Eriksson}, K. 2013, \aap,
  552, A20

\bibitem[{{Nowotny} {et~al.}(2011){Nowotny}, {Aringer}, {H{\"o}fner}, \&
  {Lederer}}]{NoAHL11}
{Nowotny}, W., {Aringer}, B., {H{\"o}fner}, S., \& {Lederer}, M.~T. 2011, \aap,
  529, A129

\bibitem[{{Nowotny} {et~al.}(2010){Nowotny}, {H{\"o}fner}, \&
  {Aringer}}]{NowHA10}
{Nowotny}, W., {H{\"o}fner}, S., \& {Aringer}, B. 2010, \aap, 514, A35

\bibitem[{{Parada} {et~al.}(2021){Parada}, {Heyl}, {Richer}, {Ripoche}, \&
  {Rousseau-Nepton}}]{Parada+21}
{Parada}, J., {Heyl}, J., {Richer}, H., {Ripoche}, P., \& {Rousseau-Nepton}, L.
  2021, \mnras, 501, 933

\bibitem[{{Pastorelli} {et~al.}(2020){Pastorelli}, {Marigo}, {Girardi},
  {Aringer}, {Chen}, {Rubele}, {Trabucchi}, {Bladh}, {Boyer}, {Bressan},
  {Dalcanton}, {Groenewegen}, {Lebzelter}, {Mowlavi}, {Chubb}, {Cioni}, {de
  Grijs}, {Ivanov}, {Nanni}, {van Loon}, \& {Zaggia}}]{Pastorelli+20}
{Pastorelli}, G., {Marigo}, P., {Girardi}, L., {et~al.} 2020, \mnras, 498, 3283

\bibitem[{{Pastorelli} {et~al.}(2019){Pastorelli}, {Marigo}, {Girardi}, {Chen},
  {Rubele}, {Trabucchi}, {Aringer}, {Bladh}, {Bressan}, {Montalb{\'a}n},
  {Boyer}, {Dalcanton}, {Eriksson}, {Groenewegen}, {H{\"o}fner}, {Lebzelter},
  {Nanni}, {Rosenfield}, {Wood}, \& {Cioni}}]{Pastorelli+19}
{Pastorelli}, G., {Marigo}, P., {Girardi}, L., {et~al.} 2019, \mnras, 485, 5666

\bibitem[{{Querci} {et~al.}(1974){Querci}, {Querci}, \& {Tsuji}}]{Quer+74}
{Querci}, F., {Querci}, M., \& {Tsuji}, T. 1974, \aap, 31, 265

\bibitem[{{Ramstedt} \& {Olofsson}(2014)}]{RO14}
{Ramstedt}, S. \& {Olofsson}, H. 2014, \aap, 566, A145

\bibitem[{{Richer} {et~al.}(1984){Richer}, {Crabtree}, \&
  {Pritchet}}]{Richer+84}
{Richer}, H.~B., {Crabtree}, D.~R., \& {Pritchet}, C.~J. 1984, \apj, 287, 138

\bibitem[{{Ripoche} {et~al.}(2020){Ripoche}, {Heyl}, {Parada}, \&
  {Richer}}]{Ripoche+20}
{Ripoche}, P., {Heyl}, J., {Parada}, J., \& {Richer}, H. 2020, \mnras, 495,
  2858

\bibitem[{{Rouleau} \& {Martin}(1991)}]{RM91}
{Rouleau}, F. \& {Martin}, P.~G. 1991, \apj, 377, 526

\bibitem[{{Sacuto} {et~al.}(2011){Sacuto}, {Aringer}, {Hron}, {Nowotny},
  {Paladini}, {Verhoelst}, \& {H{\"o}fner}}]{Sacu+11}
{Sacuto}, S., {Aringer}, B., {Hron}, J., {et~al.} 2011, \aap, 525, A42

\bibitem[{{Sch{\"o}ier} \& {Olofsson}(2001)}]{SO01}
{Sch{\"o}ier}, F.~L. \& {Olofsson}, H. 2001, \aap, 368, 969

\bibitem[{{Soszy{\'n}ski} {et~al.}(2009){Soszy{\'n}ski}, {Udalski},
  {Szyma{\'n}ski}, {Kubiak}, {Pietrzy{\'n}ski}, {Wyrzykowski}, {Szewczyk},
  {Ulaczyk}, \& {Poleski}}]{Sosz+09}
{Soszy{\'n}ski}, I., {Udalski}, A., {Szyma{\'n}ski}, M.~K., {et~al.} 2009,
  \actaa, 59, 239

\bibitem[{{Srinivasan} {et~al.}(2016){Srinivasan}, {Boyer}, {Kemper},
  {Meixner}, {Sargent}, \& {Riebel}}]{Srinivasan+16}
{Srinivasan}, S., {Boyer}, M.~L., {Kemper}, F., {et~al.} 2016, \mnras, 457,
  2814

\bibitem[{{Totten} {et~al.}(2000){Totten}, {Irwin}, \& {Whitelock}}]{Tott+00}
{Totten}, E.~J., {Irwin}, M.~J., \& {Whitelock}, P.~A. 2000, \mnras, 314, 630

\bibitem[{{Weinberg} \& {Nikolaev}(2001)}]{WN01}
{Weinberg}, M.~D. \& {Nikolaev}, S. 2001, \apj, 548, 712

\bibitem[{{Whitelock} {et~al.}(2006){Whitelock}, {Feast}, {Marang}, \&
  {Groenewegen}}]{W+06}
{Whitelock}, P.~A., {Feast}, M.~W., {Marang}, F., \& {Groenewegen}, M.~A.~T.
  2006, \mnras, 369, 751

\bibitem[{{Wittkowski} {et~al.}(2017){Wittkowski}, {Hofmann}, {H{\"o}fner}, {Le
  Bouquin}, {Nowotny}, {Paladini}, {Young}, {Berger}, {Brunner}, {de
  Gregorio-Monsalvo}, {Eriksson}, {Hron}, {Humphreys}, {Lindqvist}, {Maercker},
  {Mohamed}, {Olofsson}, {Ramstedt}, \& {Weigelt}}]{Witt+17}
{Wittkowski}, M., {Hofmann}, K.~H., {H{\"o}fner}, S., {et~al.} 2017, \aap, 601,
  A3

\bibitem[{{Wood} {et~al.}(1999){Wood}, {Alcock}, {Allsman}, {Alves}, {Axelrod},
  {Becker}, {Bennett}, {Cook}, {Drake}, {Freeman}, {Griest}, {King}, {Lehner},
  {Marshall}, {Minniti}, {Peterson}, {Pratt}, {Quinn}, {Stubbs}, {Sutherland},
  {Tomaney}, {Vandehei}, \& {Welch}}]{Wood99}
{Wood}, P.~R., {Alcock}, C., {Allsman}, R.~A., {et~al.} 1999, in IAU Symposium,
  Vol. 191, Asymptotic Giant Branch Stars, ed. T.~{Le Bertre}, A.~{Lebre}, \&
  C.~{Waelkens}, 151

\bibitem[{{Yurchenko} {et~al.}(2018){Yurchenko}, {Szab{\'o}}, {Pyatenko}, \&
  {Tennyson}}]{Yurch+18}
{Yurchenko}, S.~N., {Szab{\'o}}, I., {Pyatenko}, E., \& {Tennyson}, J. 2018,
  \mnras, 480, 3397

\end{thebibliography}
%

\begin{appendix}
\section{Details on selected cases}
\label{App_A}

\begin{figure}
  \resizebox{0.99\hsize}{!}{\includegraphics{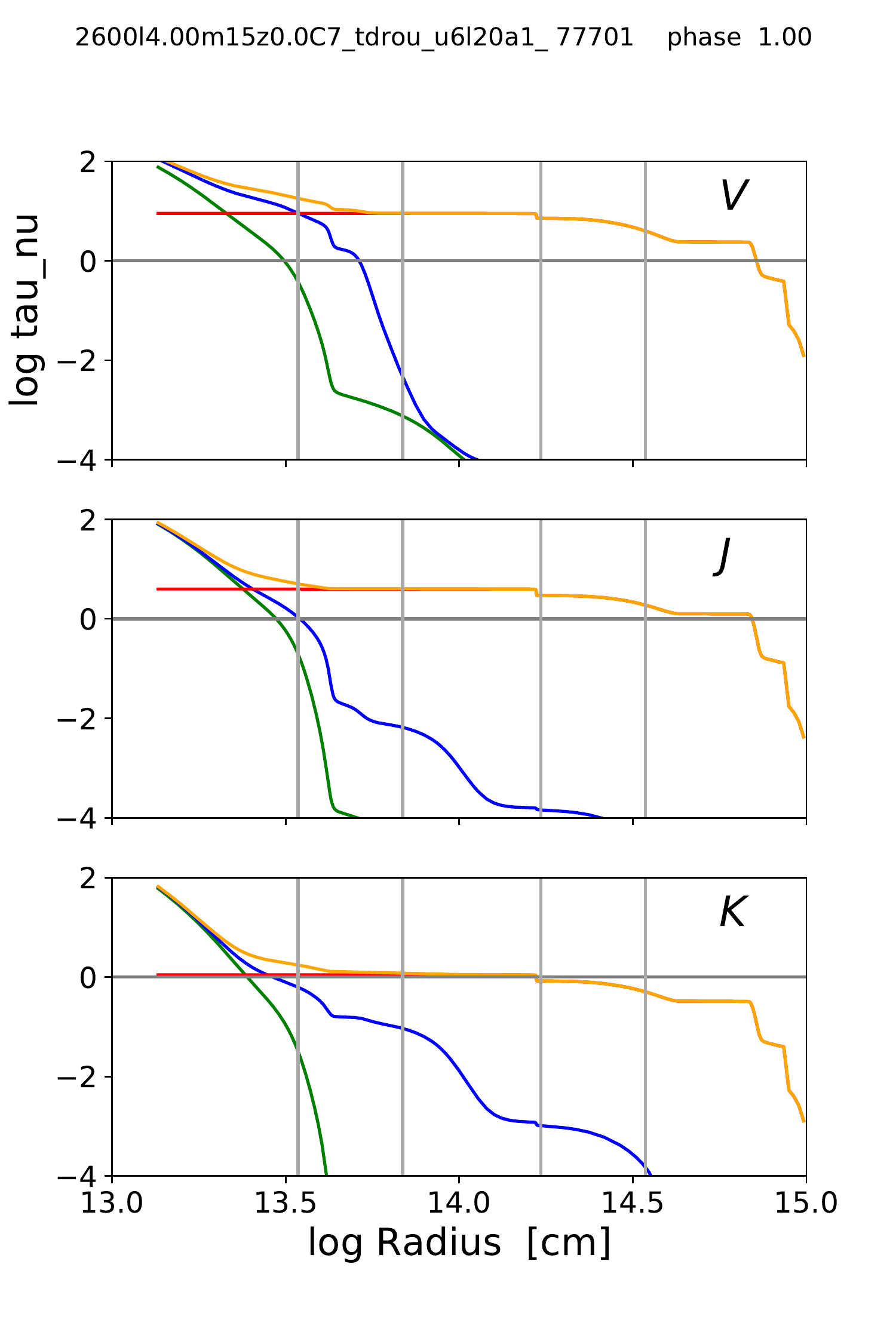}}   
  \caption{log optical depth vs log distance to stellar centre: green denote continuous opacities, blue line opacities,
  red dust opacities, and orange total opacity; vertical lines at 1, 2, 5, 10 stellar radii. Model is 2600/4.0/1.5
  C7u6. SDDO and Max phase. }
  \label{App_tau_C7u6_sddo_10}
\end{figure}

\begin{figure}
  \resizebox{0.99\hsize}{!}{\includegraphics{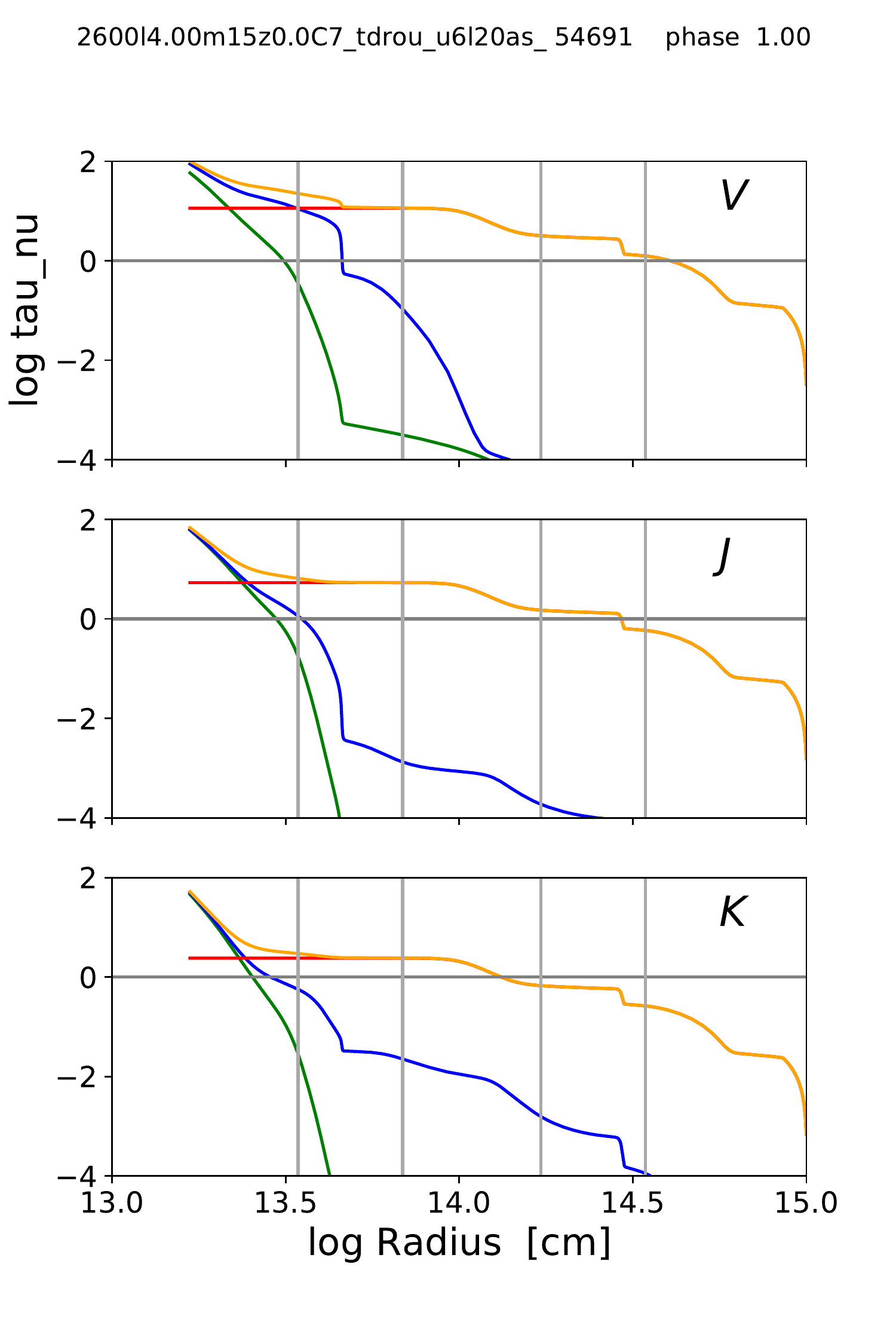}}   
  \caption{Log optical depth versus log distance to stellar centre: green continuous opacities, blue line opacities,
  red dust opacities, and orange total opacity; vertical lines at 1, 2, 5, 10 stellar radii. Model is 2600/4.0/1.5
  C7u6. SPL and max phase. }
  \label{App_tau_C7u6_spl_10}
\end{figure}

For the model 2600/4.0/1.5 with C7 and u6, we plot the optical depths as a function of radial distance, which is seen at
a typical wavelength in the $V$, $J$, and $K$ bands. The different curves correspond to continuous, line, and
dust opacities. In Fig.~\ref{App_tau_C7u6_sddo_10}, the values for the SDDO case are shown, and those for
the SPL case are given in Fig.~\ref{App_tau_C7u6_spl_10}.

\begin{figure}
  \resizebox{1.1\hsize}{!}{\includegraphics{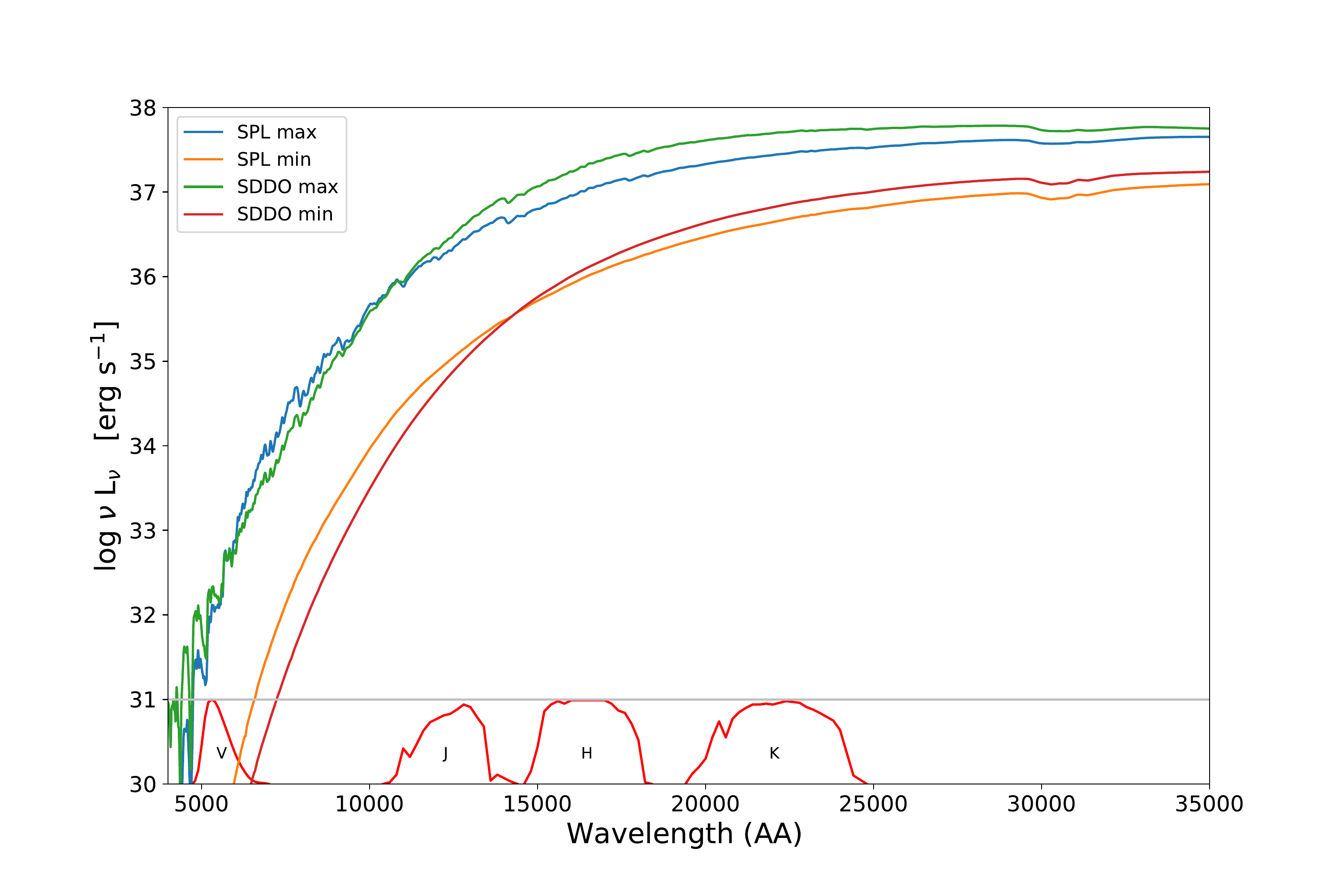}}   
  \caption{Spectra for the model 2600/4.0/1.5 C7u6 at a maximum and a minimum phase, as well as for the SPL and SDDO cases
(see legend).}
  \label{spe_26400m15C7u6}
\end{figure}

Spectra for the same model are shown in Fig.~\ref{spe_26400m15C7u6} for a maximum and a minimum phase and for both the SPL and SDDO
cases. The strong influence of dust opacity is clearly seen. The quantitative difference between SPL and SDDO is mainly due to
the decreased carbon condensation degree in the SDDO case (see Fig.~\ref{figDMMa1s}).
For the SDDO case we see a brighter $H$ magnitude at the maximum phase and a fainter $J$ magnitude at minimum,
both giving a larger ($J$-$H$) colour; this is also contributing to the steeper slope for SDDO cases in Fig.~\ref{figDMMa1asJHHK}.

\begin{figure}
  \resizebox{1.1\hsize}{!}{\includegraphics{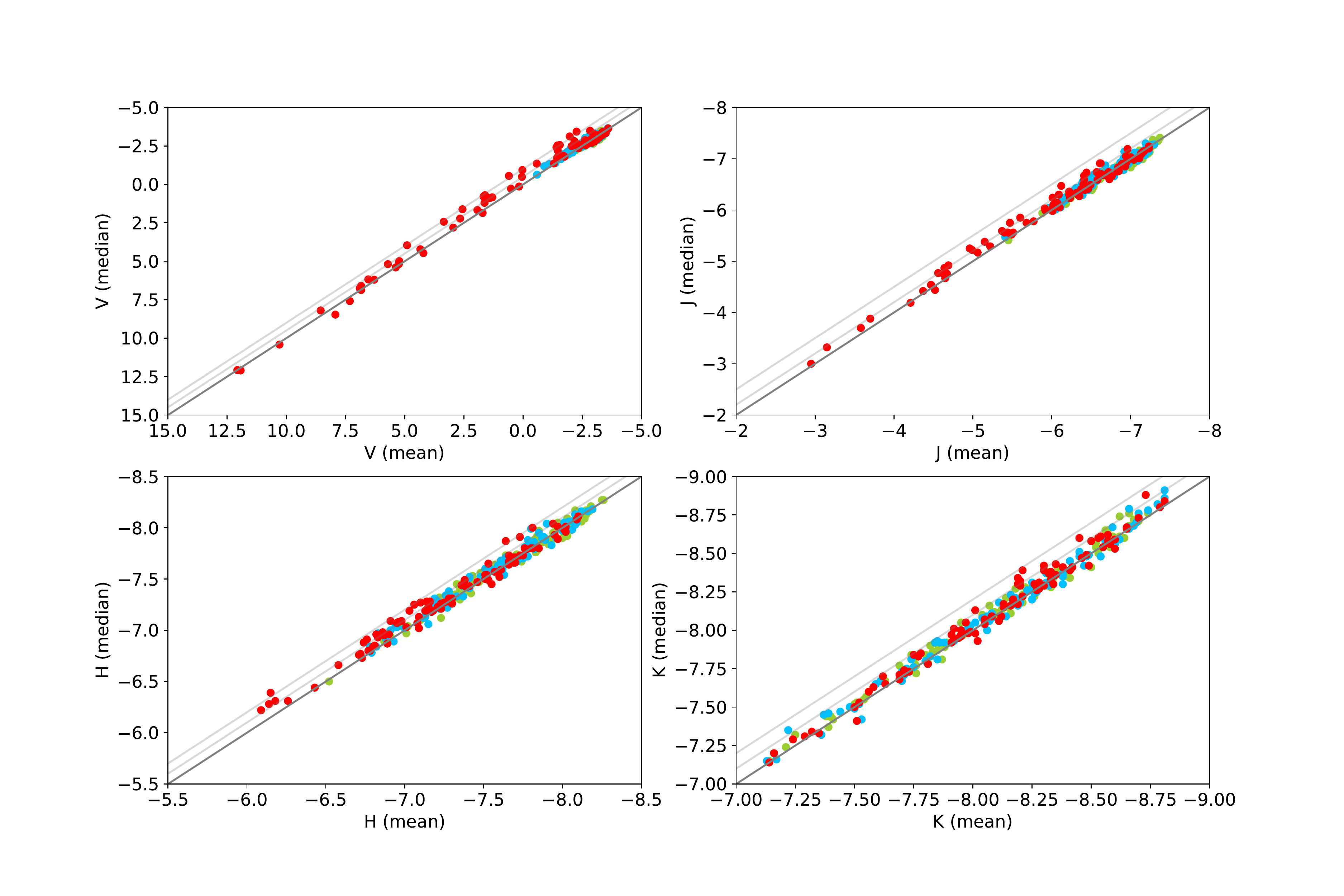}}   
  \caption{Median magnitudes versus mean magnitudes plotted for models in the SDDO grid.
  The different panels show the effects in the $V$, $J$, $H$, and $K$ filters. The models are
  colour-coded according to their carbon excess (see Fig.~\ref{figDMMa1asKmagndus}). Offsets from the 
  one-to-one line of 0.5 and 1.0 (in $V$), 0.2 and 0.5 (in $J$), and 0.1 and 0.2 (in $H$ and $K$) are
  indicated by the grey lines.
  }
  \label{medianmeanVJHK}
\end{figure}

\section{Median versus mean magnitudes}
\label{a:medianmean}
If a light curve is asymmetric in time, for example with rather narrow minima and broader maxima,
there can be a difference between the mean and the median magnitude.
We also calculated the median for the light curves of the models in the SDDO grid.
In Fig.~\ref{medianmeanVJHK}, we show the effects in the $V$, $J$, $H$, and $K$ filters.
We note that the effect of replacing the mean by the median magnitude results in
generally brighter magnitudes, but the effect is small, except for some of the highest
carbon excess (C7) models.

\section{Model overview}
\label{a:overviewdata}

A table with photometric and dynamic properties of
the models in the present grid is found in electronic form at CDS via anonymous ftp to 
cdsarc.cds.unistra.fr (130.79.128.5)
\footnote{or via https://cdsarc.cds.unistra.fr/cgi-bin/qcat?J/A+A/}.
The models are arranged by increasing effective temperature, luminosity, and stellar mass.
For each such combination, the data are ordered by increasing carbon excess and piston velocity
amplitude.
In each line, after the model parameters we list the log~g (surface gravity in cgs units).
Then come dynamic quantities evaluated at the outer boundary:  mass-loss rate
(in solar masses per year),
the wind velocity (km/s), the carbon condensation degree, and the dust-to-gas ratio.
We note that all given values are temporal means (see Sect.~\ref{s:DMAdyn}).
Then follow the photometric properties: the (full)
amplitude of the bolometric magnitude, the mean $V$ magnitude, the range of $V$ magnitudes, the mean
$K$ magnitude and its range, and finally the colours \mbox{($V$\,--\,$I$)}, \mbox{($V$\,--\,$K$)},
\mbox{($J$\,--\,$H$),} and \mbox{($H$\,--\,$K$)}.

The luminosities and stellar masses are given in solar units. 
The carbon excesses, log(C--O)+12,
 are given on the scale where 
log N$_H$ $\equiv$12.00. 
The piston velocity amplitudes, $\Delta u_{\rm p}$, are given in km/s. 
All the  photometric quantities are given in magnitudes.
The first  lines are shown below as an example.

\onecolumn

\begin{landscape}

\small

\begin{longtable}{cccccccccccrrrrrrcc}
\hline\hline
\bigstrut[t]
$T_\star$ & log\,$L_\star$ & $M_\star$ & log(C--O)+12 & $\Delta u_{\rm p}$ & log g & log $\dot{M}$
&  $u_\infty$  & $f_c$  & $\rho_d$/$\rho_g$ & $\Delta m_{bol}$ & $V$ & $\Delta V$ &  $K$ &  $\Delta K$ 
& \mbox{($V$\,--\,$I$)} & \mbox{($V$\,--\,$K$)} & \mbox{($J$\,--\,$H$)} &   \mbox{($H$\,--\,$K$)}  \bigstrut[b] \\
\hline
 2600 & 3.70 & 0.75 & 8.2 &  2 & -0.79 & --     & --   & --    & --       & 0.34 & -2.28 &  0.83 & -7.89 & 0.43 & 2.71 &  5.61 & 0.83 & 0.51 \bigstrut[t] \\ 
 2600 & 3.70 & 0.75 & 8.2 &  4 & -0.79 & --     & --   & --    & --       & 0.68 & -2.19 &  1.82 & -7.85 & 0.94 & 2.75 &  5.66 & 0.79 & 0.54  \\ 
 2600 & 3.70 & 0.75 & 8.2 &  6 & -0.79 &  -8.05 &  2.8 & 0.050 & 7.02e-05 & 1.05 & -2.25 &  2.28 & -7.88 & 1.10 & 2.58 &  5.63 & 0.80 & 0.61  \\ 
 2600 & 3.70 & 0.75 & 8.5 &  2 & -0.79 &  -8.37 &  2.5 & 0.033 & 9.33e-05 & 0.34 & -2.40 &  0.71 & -7.80 & 0.45 & 2.53 &  5.40 & 0.79 & 0.56  \\ 
 2600 & 3.70 & 0.75 & 8.5 &  4 & -0.79 &  -6.20 & 12.0 & 0.059 & 1.66e-04 & 0.70 & -1.90 &  2.59 & -7.95 & 0.78 & 2.71 &  6.06 & 0.92 & 0.65  \\ 
 2600 & 3.70 & 0.75 & 8.5 &  6 & -0.79 &  -6.01 & 11.7 & 0.063 & 1.78e-04 & 1.07 & -1.77 &  3.86 & -7.97 & 1.06 & 2.74 &  6.20 & 0.93 & 0.68  \\ 
 2600 & 3.70 & 0.75 & 8.8 &  2 & -0.79 &  -6.13 & 27.1 & 0.140 & 7.86e-04 & 0.35 & -1.29 &  2.49 & -7.98 & 0.55 & 3.12 &  6.69 & 0.97 & 0.71  \\ 
 2600 & 3.70 & 0.75 & 8.8 &  4 & -0.79 &  -5.69 & 22.4 & 0.173 & 9.74e-04 & 0.71 &  1.63 & 10.45 & -8.08 & 0.94 & 4.30 &  9.70 & 1.41 & 0.99  \\ 
 2600 & 3.70 & 0.75 & 8.8 &  6 & -0.79 &  -5.40 & 19.9 & 0.233 & 1.30e-03 & 1.10 &  5.23 & 12.78 & -7.91 & 1.62 & 5.75 & 13.14 & 1.91 & 1.33  \bigstrut[b]\\ 
\hline

\caption{Dynamic and photometric properties of the models in the present grid. The meaning and units for the
different columns are described in Sect.~\ref{a:overviewdata}}

\end{longtable}

\end{landscape}

\end{appendix}

\end{document}